\pdfoutput=1
\documentclass[11pt]{article}
\usepackage[margin=1in]{geometry}
\usepackage{graphicx,booktabs,amsmath,amssymb,hyperref,natbib,longtable}
\usepackage{cmap}
\usepackage[T1]{fontenc}
\input{glyphtounicode}
\graphicspath{{./}{../results/}}
\hypersetup{colorlinks=true,linkcolor=blue,citecolor=blue,urlcolor=blue,
  pdftitle={Chemistry-resolved disagreement between Materials Project-compatible PBE(+U) and r2SCAN stability verdicts in the GNoME materials database},
  pdfauthor={Karthikeya Aduru}}
\title{\textbf{Chemistry-resolved disagreement between Materials Project-compatible
PBE(+$U$) and r$^2$SCAN stability verdicts in the GNoME materials database}}
\author{Karthikeya Aduru \\ \small Accretion Labs \\ \small \texttt{founder@accretionlabs.net}}
\date{October 2026}

\begin{document}
\sloppy 
\maketitle

\begin{abstract}
GNoME reports hundreds of thousands of thermodynamically stable materials, but its
labels rest on a Materials Project-compatible generalised-gradient-approximation
(GGA/GGA+$U$) workflow, not on a functional alone. Its authors report that 84\% of its discovered binaries
and ternaries retain a negative decomposition energy under r$^2$SCAN, and 86.8\% of
tested quaternaries remain stable on the r$^2$SCAN hull. We resolve that aggregate by chemistry over the
139{,}541 GNoME materials released with r$^2$SCAN recomputations, a subset the dataset
authors selected rather than sampled, using cluster bootstrap intervals over 8{,}429
structural-prototype proxy clusters.

The aggregate conceals a 26.7-point spread at fixed element count, from 4.0\%
[3.3, 4.7] for plain quaternary intermetallics (no lanthanide, no 3$d$ metal) to 30.7\%
[27.7, 33.8] for lanthanide-free 3$d$-containing quaternary oxides, a pair identified
post hoc against a prospectively
fixed threshold. Under our reading of the release's ``All'' reference set, the two cells sit at
14.5\% and 37.4\%, preserving a 22.9-point spread while
substantially reducing the ratio. Presence of a Materials Project Hubbard-$U$-list element is strongly associated with
disagreement, conditional on the other covariates, in the fitted model over all
139{,}541 materials
(cluster-robust odds ratio 3.45 [3.13, 3.80]). Neither selective $U$ application nor hull proximity explains away this
association under the analyses performed: disagreement stays elevated outside the
$U$-treated regime, and a natural cubic spline in the continuous PBE decomposition energy
moves the odds ratio in the 117{,}018-material model only from 3.53 to 3.43 [3.07, 3.83]. For allocating a
verification budget, composition supplements hull proximity rather than replacing it: over
held-out cluster splits at a 20\% budget, the two together surface 45.8\% of
disagreements against 40.4\% for hull distance alone. An exploratory
comparison against previously catalogued materials changes direction between populations.
We measure
disagreement between two workflows, not accuracy.
\end{abstract}

\section{Introduction}

Computational materials discovery has shifted from enumerating candidates to generating
them. GNoME (Graph Networks for Materials Exploration) reported 381{,}000 new entries on the
convex hull, the surface of lowest formation energy below which a composition has no
cheaper decomposition into competing phases, for a total of 421{,}000
stable crystals \citep{merchant2023}; subsequent releases, which widened the
release criterion from the documented on-hull tolerance of $5\times10^{-5}$ eV/atom to
within 1 meV/atom of the hull \citep{gnomedataset}, brought the public summary file to
554{,}054 entries (our row count). Generative models such as MatterGen
\citep{zeni2025} and expanded databases such as Alexandria have followed, the latter's
PBEsol/SCAN subset \citep{schmidt2022} using PBEsol geometries with SCAN energies
rather than PBE. GNoME's own stability verdicts, which are
what this paper analyses, rest on density functional theory (DFT) calculations at the
Perdew--Burke--Ernzerhof (PBE) level \citep{pbe1996}, executed in VASP \citep{vasp1996}
with Materials Project (MP)-compatible settings \citep{jain2013}, including a Hubbard $U$
for eight transition metals in oxides and fluorides and the Materials Project 2020 (MP2020) correction scheme's empirical anion and
GGA/GGA+$U$ mixing corrections \citep{wang2021mp2020}. The functional and
correction choices differ across MatterGen, Alexandria and GNoME, so the question we ask
of GNoME does not transfer to the others unchanged.

GGA formation energies carry errors of order 0.1 eV/atom \citep{kirklin2015}, but
formation-energy errors do not directly determine stability-verdict disagreement,
because a decomposition energy compares a material against composition-conserving
combinations of competing phases rather than against an elemental reference. We
therefore analyse the released decomposition-energy outcomes directly. The need for
chemistry resolution is practical: a laboratory choosing which predicted compound to attempt cannot act on
an aggregate error rate, a database maintainer deciding where to spend a finite
recomputation budget faces the same problem at scale, and aggregate benchmark scores
against PBE-derived labels \citep{riebesell2025} do not show how sensitivity to the
workflow varies across chemistry.

GNoME's authors anticipated the question. They recomputed a subset with the r$^2$SCAN
meta-GGA, the regularised-restored form of the SCAN functional \citep{furness2020}, and reported that 84\% of discovered binaries and ternaries
retain negative phase-separation energies (their term for the decomposition energy), with
86.8\% of tested quaternaries remaining
stable on the r$^2$SCAN hull \citep{merchant2023}. Supplementary Note 2 of that paper observes that the
discrepancy concentrates near the hull and that the same effect appears in the Materials
Project and the Open Quantum Materials Database (OQMD), but gives no breakdown by chemistry. The authors released the underlying
r$^2$SCAN data with the statement that it is provided ``expectantly serving as a foundation
for analysing discrepancies between functional choices.''

\subsection{What this paper measures, and what it does not}\label{sec:scope}

We measure \emph{disagreement between two computational methods}: a material released as
stable at the PBE level whose r$^2$SCAN decomposition energy is positive (defined precisely
in Section~\ref{sec:defs}). We do not establish
which functional is closer to the truth for any material, and we say nothing about
whether these compounds are stable in reality. We therefore use the term \emph{flip}
throughout and avoid calling flipped labels ``incorrect''. Where we discuss practical
consequences the operative quantity is the share of labels a higher-fidelity
recomputation would revise, which is a statement about what the expensive calculation
would say, not about nature. This constrains every claim that follows, including the
budget model of Section~\ref{sec:budget}, whose payoff is measured in disagreements
surfaced rather than errors corrected.

\subsection{Questions and contributions}\label{sec:contrib}

\begin{enumerate}
\item \textbf{Q1.} Does the rate of disagreement vary systematically with chemistry,
controlling for element count, and which compositional features is it associated with?
\item \textbf{Q2} (exploratory). Do GNoME materials flip at a different rate than
previously catalogued materials in matched chemistry? We label this exploratory in the
question list, not only in its results section, because the two populations differ in
provenance, selection and phase-diagram completeness as well as in chemistry.
\item \textbf{Q3.} Given a fixed recomputation budget, does composition add information
about disagreement beyond proximity to the PBE hull, and by how much?
\end{enumerate}

Our contributions are: (i) a chemistry- and element-resolved map of stability-verdict
disagreement between the two workflows on GNoME; (ii) stratified comparisons showing that a lanthanide-pooled form of the composition
contrast is present within every populated hull-distance band, that the 3$d$-associated excess is
present within every band-gap bin, and that the contrast persists with cell size as a
covariate, under inverse-probability weighting (IPW) for recomputation selection
(Section~\ref{sec:ipw}), and outside the Hubbard-$U$-treated regime; (iii) the observation that the binary verdict change
reflects a chemistry-dependent sign in the underlying energy displacement; (iv) an
exploratory matched comparison against previously catalogued materials whose direction
varies with composition; (v) a verification-budget evaluation over repeated cluster-wise
splits showing that composition supplements rather than replaces proximity to the hull;
and (vi) the observation that the compositional regime GNoME advertises most prominently
is recomputed at the meta-GGA level too sparsely for stratified analysis.

Decision rules, including the threshold that would have caused us to abandon the
chemistry-resolved framing, were committed to version control before the stratified
analysis was run, following an earlier unstratified inspection of the data that informed
the secondary lanthanide hypothesis (Section~\ref{sec:prereg}). We write
``pre-committed'' rather than ``pre-registered'' throughout: the rules were prospectively
committed to a public repository, but a repository commit is not an independently
timestamped registration.

\section{Related work}

\paragraph{Functional comparisons on known materials.}\label{sec:related-func}
Kingsbury et al.\ benchmarked r$^2$SCAN against SCAN and PBEsol across roughly 6{,}000
solids and found r$^2$SCAN more accurate than either for formation energies
\citep{kingsbury2022prm}, and developed the mixing scheme the Materials Project uses to
place PBE(+$U$) and r$^2$SCAN energies on a single phase diagram \citep{kingsbury2022npj}.
Two results bear directly on what follows. Kothakonda et al.\ tested r$^2$SCAN
thermochemistry on more than 1{,}000 solids and concluded that for transition metals,
intermetallics and weakly bound solids, GGAs are comparable to meta-GGAs, and that for
the reaction class they label Type 2 decomposition enthalpies, PBE is slightly more accurate \citep{kothakonda2023}.
Bartel et al.\ reached a compatible verdict from the reaction side: across 646
decomposition reactions that are not simply formation reactions, PBE (mean absolute error 70 meV/atom) and
SCAN (59 meV/atom) perform similarly \citep{bartel2019}. Both are statements about error
magnitude against experiment, not about agreement between the two functionals:
Section~\ref{sec:shift} shows that intermetallics flip rarely while their mean
displacement is negative, which is consistent with a sign effect rather than with the
functionals agreeing.

\paragraph{Correction schemes, chemistry-resolved error, and critiques of the release.}
The MP2020 scheme quantifies uncertainty in DFT energy corrections and shows that many
polymorphs classified as unstable may be stable once that uncertainty is propagated
\citep{wang2021mp2020}. Warford, Thiemann and Cs\'anyi argue that applying $U$ selectively
--- only when oxygen or fluorine is present --- creates two mutually incompatible
potential-energy surfaces within one dataset \citep{warford2026}; Section~\ref{sec:u}
examines the association across that boundary. Sun et al.\ data-mined 29{,}902 observed
phases to produce a chemistry-resolved scale of metastability \citep{sun2016}, and Yu et
al.\ mapped universal machine-learned interatomic potential (MLIP) formation-energy error per element across 19{,}998 unary and
binary phases \citep{yu2024}; both stratify a computed-energy error by chemistry, and
neither addresses functional-level verdict flips in a generated candidate database.
Cheetham and Seshadri question how many GNoME entries are simultaneously novel,
credible and useful \citep{cheetham2024}, Liu and Meng report that metals constitute
over 60\% of GNoME \citep{liu2024}, and Leeman et al.\ examine the companion
autonomous-synthesis work \citep{leeman2024,szymanski2023,szymanski2026}. None of these
addresses label reliability.

\paragraph{Flip-rate framing.}
The closest prior framing is Reents, Bercx and Pizzi, whose supplementary material plots a
PBEsol-versus-SCAN stability flip rate against threshold for Alexandria
\citep{reents2026}. That analysis is aggregate, concerns a different database and
functional pair, serves to validate a machine-learned correction, and reports no
chemistry resolution. An et al.\ re-screened a machine-learning-generated set, finding 298 of 603
curated structures remained PBE-stable on complete hulls and 109 of 166 dynamically
stable candidates survived a SCAN refinement \citep{an2026}; they excluded 142 rare-earth
compounds outright because of $f$-electron and pseudopotential uncertainty. Huang et al.\
map PBE formation energies onto the r$^2$SCAN scale across 380{,}190 paired MatPES
entries \citep{huang2026,kaplan2025} but verify stability verdicts for only six binary
systems. We are not aware of a prior chemistry- or element-resolved verdict-flip analysis
on GNoME or any comparable generated database, and state this as the result of our own
literature search rather than as an established priority claim.

\section{Data and methods}

\subsection{Datasets}\label{sec:datasets}

\paragraph{GNoME.} We use two files from the public release \citep{merchant2023}. The summary
file, \texttt{stable\_materials\_summary.csv}, holds 554{,}054 unique material IDs with
PBE-level decomposition energies, band gaps, space groups and cell information. The
recomputation file, \texttt{stable\_materials\_r2scan.csv}, holds 139{,}541 unique IDs
with r$^2$SCAN recomputations. It carries composition, element list, site count, space
group and crystal system of its own, so the prototype clustering of
Section~\ref{sec:cluster} and the log-sites term of Section~\ref{sec:logit} are defined
for all 139{,}541. Band gap and the PBE decomposition energy come only from the summary
file, so any analysis using those is restricted to the 117{,}043 materials present in
both, and we state the restricted $n$ wherever it applies. Positive r$^2$SCAN
decomposition energies are present throughout the recomputation file, so it is not
censored to survivors.

Each file reports decomposition energy against several reference sets, which the release
documents as distances to four convex hulls: \emph{MP}, the hull of all Materials Project
entries; \emph{MP OQMD} (written MP+OQMD below), Materials Project plus OQMD entries;
\emph{All}, the hull of all entries (the documentation's convex-hull section says that
hull is built from Materials Project, OQMD and WBM (Wang--Botti--Marques dataset) energies),
which we read as also including GNoME's other structures; and \emph{Relative}, the hull of all entries except the current one \citep{gnomedataset}.
That documentation is written for the summary file; we take the documented meanings as
applying to the recomputation file's identically named columns. The released
All-referenced r$^2$SCAN energies are positive for 25.8\% of entries
(Section~\ref{sec:allref}), which could not occur if each entry's own energy were part
of its reference hull, so we read the All column as excluding the entry itself, which on that reading makes
it the leave-one-out quantity the \emph{Relative} column was meant to hold; that column,
which would otherwise be the most natural outcome, is empty in both released files. We take the MP-referenced
column as primary because it reproduces the published aggregate, report the MP+OQMD
variant as a sensitivity check, and examine the ``All'' column separately in
Section~\ref{sec:allref} because it gives a materially different result. The release does not
document how the MP+OQMD and All hulls were built for the recomputation file; OQMD and WBM
carry no r$^2$SCAN energies, so those hulls may mix functionals, and we treat rates against
them as sensitivity checks rather than as hulls of known construction.
Rates against ``All'' are computed from the recomputation file.

Of the 117{,}043 recomputed materials that can be matched to a PBE decomposition energy,
25 have a positive PBE value (the current release admits entries up to 1 meV/atom above
the hull) and so are not PBE-stable under our $\le 0$ convention. They cannot be
confirmed transitions and are excluded from every analysis that uses PBE energies; they are 0.02\% of the set
and no reported number changes. ``PBE-stable'' throughout means a non-positive MP-referenced
PBE decomposition energy in the summary file, a hull of Materials Project entries; it is not
the GNoME-inclusive hull on which the release's own stability claim rests.

\paragraph{How the r$^2$SCAN energies were produced.} Each material passes through a
PBE pre-relaxation with PBE54 potentials, an r$^2$SCAN relaxation and an r$^2$SCAN
static calculation at \texttt{MPScanRelaxSet} and \texttt{MPScanStaticSet} settings
\citep{merchant2023}, so the structures are re-relaxed rather than evaluated at the PBE
geometry, the projector-augmented-wave (PAW) datasets can differ slightly from those behind the PBE labels, and
every calculation is initialised ferromagnetically. A verdict change therefore reflects
the functional, the correction scheme, the PAW dataset, the relaxed geometry and
possibly the converged magnetic state, together. The release does not state how the
recomputed subset was chosen, nor whether any correction is applied to the r$^2$SCAN
energies.

\paragraph{Known materials.} We pull Materials Project thermo records separately for the
\texttt{GGA\_GGA+U} and \texttt{r2SCAN} thermo types and keep entries present in both
(Appendix~\ref{app:manifest} gives the exact query and a spelling pitfall in the thermo
type). We retrieved 37{,}123 r$^2$SCAN and 152{,}426 GGA+$U$ documents; 35{,}176 material
IDs carry both, of which 22{,}756 are stable at the GGA(+$U$) level and therefore capable of
flipping. The \texttt{r2SCAN} thermo type is distinct from the Materials Project's mixed
scheme \citep{mpmixingdocs}: its hull is built from whichever r$^2$SCAN entries exist in
a chemical system, with no requirement that every competing phase has been recomputed
(Section~\ref{sec:allref}).

\subsection{What \texorpdfstring{``PBE''}{"PBE"} means here}

Throughout, ``PBE'' denotes the Materials Project-compatible level at which GNoME's
labels were produced: PBE with, for eight transition metals (V, Cr, Mn, Fe, Co, Ni, Mo
and W; the ``$U$ list'' below) in oxides and fluorides, a Hubbard $U$ applied in the
calculation, and the MP2020 scheme --- implemented in
the pymatgen library as \texttt{MaterialsProject2020Compatibility} --- which adds empirical anion
corrections and a per-atom GGA/GGA+$U$ mixing adjustment for those same elements. It is
not bare PBE. We keep the short name because it is the
one the source work and the community use, but the difference between the corrected and
uncorrected surfaces is the subject of Section~\ref{sec:u} and should be kept in mind
wherever this paper says ``PBE''.

\subsection{Population flow}\label{sec:popflow}

Different analyses run on different subsets, because the required fields are not
available for every material. Table~\ref{tab:popflow} states each population once so that
the varying $n$ throughout the paper can be traced to a stated exclusion.

\begin{table}[htbp]\centering
\caption{Populations used in this paper, with the exclusion that defines each.}
\label{tab:popflow}
\small
\begin{tabular}{lrl}
\toprule
Population & $n$ & Defined by \\
\midrule
Summary release                  & 554{,}054 & all entries in the PBE summary file \\
\quad PBE-stable, $\le$4 elements & 365{,}504 & decomposition energy $\le 0$, $n_{\text{el}}\le4$ \\
Recomputation release            & 139{,}541 & all entries in the r$^2$SCAN file \\
\quad matched to the summary file & 117{,}043 & \texttt{MaterialId} present in both \\
\quad matched and PBE-stable      & 117{,}018 & 25 matched entries were not PBE-stable \\
\quad matched, PBE-stable, $\le$4 elements & 116{,}789 & IPW outcome, censoring bounds (Sections~\ref{sec:ipw}, \ref{sec:selection}) \\
Primary composition cells        & 135{,}044 & 40 cells with $n\ge500$ (Section~\ref{sec:defs}); binaries, \\
                                 &           & quinaries and hydrides all fall below the threshold \\
Band-gap analysis                & 112{,}434 & finite positive gap (Section~\ref{sec:gap}) \\
Budget evaluation                & 117{,}018 & needs a PBE energy; PBE-stable \\
Inverse-probability weighting (IPW) & 371{,}870 & summary entries with $\le$4 elements \\
Materials Project comparison     & \phantom{0}22{,}756 & GGA(+$U$)-stable with r$^2$SCAN thermo \\
\bottomrule
\end{tabular}
\end{table}

The first and third rows are raw release counts; the rest are analytical. The 25 matched
entries that were not PBE-stable are outside the budget evaluation and the
energy-displacement analysis of Section~\ref{sec:shift} by construction, since both
require an observed non-positive PBE energy; the IPW rate of
Section~\ref{sec:ipw} is computed on the 116{,}789 IPW-eligible entries, which likewise
excludes them, while the 371{,}870 row is that model's \emph{fitting} population and not
an outcome population. The 25 are retained in the release-based rows, where they are 24
of the 135{,}044 primary-cell materials.

\subsection{Definitions}\label{sec:defs}

We report two outcomes for GNoME and they must be kept apart. \emph{Release-based
disagreement} is a positive r$^2$SCAN decomposition energy for an entry of the
recomputation release; it is defined for all 139{,}541 entries and makes no use of the PBE
verdict. A \emph{confirmed transition} additionally requires that the entry's PBE
decomposition energy is observed in the summary file and is non-positive; it is defined
for the 117{,}018 matched, PBE-stable entries. For the 22{,}498 entries with no summary-file row, PBE stability is \emph{inferred}
from the release's provenance and cannot be checked. Dropping the 25 observed-unstable
entries moves the release-based rate from 13.041\% to 13.040\%; the choice is
numerically immaterial, but the two outcomes are not the same quantity and we label
them where both appear. We use \emph{flip} as shorthand for release-based disagreement throughout, and
rates quoted without qualification are release-based. For Materials Project entries,
where \texttt{energy\_above\_hull} is non-negative by construction, we define stability as
$E_{\text{hull}} \le 10^{-6}$ eV/atom and flipping as GGA(+$U$)-stable with
$E_{\text{hull}}^{\text{r}^2\text{SCAN}} > 10^{-6}$.

We classify each composition into one of seven mutually exclusive anion classes by
priority (the classes are coarse with respect to the MP2020 scheme, which corrects Si
but not B or C among the boride, carbide and silicide class): oxide (contains O); halide (F, Cl, Br or I, no O); chalcogenide (S, Se or Te,
none of the above); pnictide (N, P, As, Sb or Bi, none of the above); hydride (H, none of
the above); boride/carbide/silicide (B, C or Si, none of the above); intermetallic (none
of the above). We flag the presence of a lanthanide (La--Lu; ``Ln'' in tables) and of a
3$d$ transition metal (Sc--Zn). Cells are the Cartesian product of element count, anion class, lanthanide
flag and 3$d$ flag, and are disjoint by construction. Only cells with $n \ge 500$ enter
the primary analysis: forty cells qualify, holding 135{,}044 of the 139{,}541 materials.
No cell is excluded by rule: the 4{,}497 materials outside the forty cells all sit in
cells below the threshold, namely the binaries, the quinaries, the hydride class and roughly
2{,}000 materials in other thin cells, some of which sit at the high-flipping end, so
the reported spread is if anything conservative. Per-element analyses use $n \ge 300$ and are not restricted
by element count. Secondary analyses apply further minimum counts stated where they are
used: 100 known materials per matched cell, 100 element-containing materials per class,
and 200 materials per hull-distance band.

\subsection{Materials are not independent observations}\label{sec:cluster}

GNoME is substitution-generated: the authors attribute 232{,}477 of 381{,}000 stable
structures to symmetry-aware partial substitution \citep{merchant2023}. Materials sharing
a parent structure and differing by one or two ionic substitutions are not independent
samples of chemical space, and may share relaxation behaviour, competing phases and
systematic error.

We quantify this using a structural-prototype proxy: the pair (space group number,
anonymised stoichiometry), where anonymised stoichiometry replaces element identities with
sorted site counts, so that Nd$_7$Os$_1$Pr$_3$Si$_5$ becomes A1B3C5D7. The 139{,}541
recomputed materials fall into \textbf{8{,}429 such clusters}, mean size 16.55; the five
largest hold 2{,}953 to 1{,}122 materials, and 87.5\% of all materials sit in clusters of
ten or more. The intra-cluster correlation of the flip outcome is $\rho = 0.138$. The
equal-size design-effect formula $1 + (\bar m - 1)\rho$ with $\bar m = 16.55$ gives 3.15;
the unequal-size formula $1 + (\sum_g m_g^2/n - 1)\rho$ gives 46.3, because five clusters
hold 7.7\% of the data. Neither approximation is used for
inference. Their implied standard-error multipliers are approximately 1.77 and 6.80;
the empirical interval-widening factors, from the cluster bootstrap described below and
from a cluster-robust sandwich estimator, are 1.82 for the median composition cell and
2.77 for the $U$-list coefficient.

\textbf{Every interval attached to a GNoME flip rate in this paper is a cluster bootstrap
interval}: we resample prototype clusters with replacement, not individual materials, at
20{,}000 replicates for cells and for the per-element analysis and 100{,}000 for the
within-anion excesses (Appendix~\ref{app:impl} gives the count for each analysis). One
population is not clustered this way and is labelled where it appears: the Materials
Project entries of Section~\ref{sec:known}, for which the thermo fields we retrieved carry
no structural information our clustering key could use. Across the 40 primary cells the
observed widening relative to naive intervals has median 1.82 and ranges from 1.17 to
5.16 (Appendix~\ref{app:cells}). Point estimates are unaffected; only the intervals are. The odds-ratio intervals of
Section~\ref{sec:logit} come from the cluster-robust sandwich estimator rather than a
bootstrap; its average marginal effects use 1{,}000 cluster bootstrap refits. Where a table gives point estimates without intervals it is descriptive and no
claim in the text rests on the precision of those numbers.

The prototype proxy is a choice, so we test it. Clustering on anonymised stoichiometry
alone, ignoring space group (2{,}186 clusters, mean size 63.8), leaves the headline
quaternary contrast essentially unaffected: 30.7\% [27.4, 34.0] against 4.0\%
[3.3, 4.7], compared with 30.7\% [27.7, 33.8] against 4.0\% [3.3, 4.7] under the
primary definition. Dropping the five largest clusters (10{,}730 materials, 7.7\% of the
set) moves the headline contrast from $+26.7$ to $+26.8$ points, the 20 largest to
$+27.3$ and the 50 largest to $+28.2$, so the largest clusters are if anything diluting
it.

\subsection{Multiple comparisons and regression uncertainty}

The per-element analysis performs 83 tests. We compute Benjamini--Hochberg adjusted
$p$-values for the contrast between each element's flip rate and the global rate. The
contrast is resampled jointly, so that each replicate recomputes both rates from one
drawn set of clusters (Appendix~\ref{app:impl}); with $\hat D_e$ the bootstrap
difference between the element rate and the global rate, the $p$-value is the two-sided
normal-approximation tail $2\left(1-\Phi\!\left(|\hat D_e| / \mathrm{se}(\hat
D_e)\right)\right)$, with $\mathrm{se}$ the standard deviation of the resampled
differences, and the Benjamini--Hochberg step-up is applied to the 83 resulting values.
The primary cell-level comparison and the lanthanide hypothesis were pre-committed; the
per-element analysis was not, and is treated as exploratory despite surviving
correction.

Logistic regression coefficients are fitted by iteratively reweighted least squares, with
standard errors from a cluster-robust sandwich estimator: with $r_g = y_g - \hat p_g$ the
residual vector of cluster $g$ and $s_g = X_g^{\top} r_g$ its score contribution,
$\hat V = B \left(\sum_g s_g s_g^{\top}\right) B$ with $B = (X^{\top} W X)^{-1}$ and
$W = \mathrm{diag}(\hat p(1-\hat p))$, clustering on the same structural prototypes used
for the bootstrap; no finite-cluster correction is applied in this model, since with 8{,}429 clusters
the factor is negligible (the hull-distance-adjusted fits of Table~\ref{tab:hullcont} and the
context model of Section~\ref{sec:additivity} apply it; Appendix~\ref{app:impl}). Relative to naive analytic errors this
inflates the standard error on the $U$-list coefficient by a factor of 2.77, between the
two design-effect approximations above.

\subsection{Pre-committed decision rules}\label{sec:prereg}

The pre-committed primary rule, evaluated on cells with $n \ge 500$ at fixed element
count, was: a spread of $\ge 5$ percentage points between the highest and lowest cell with
non-overlapping intervals supports a standalone analysis; a spread below 3 points, or
intervals that all overlap, folds the result into a single parameter reported elsewhere;
a spread of 3 to 5 points proceeds only if the comparison against known materials shows a
separate effect. The file specifies naive bootstrap intervals with 1{,}000 resamples; the
cluster intervals reported here are wider, and the rule's conditions hold under both. The
secondary pre-committed claim was that lanthanide-containing materials would flip at or
below the global rate at fixed element count, to be counted as refuted if they flipped
above it. They do, within each element count and pooled (11.8\% against 13.0\%), on point
estimates. We record three problems with that test rather than bank its result. The comparator was badly
chosen: lanthanide-containing materials are 59.4\% of the recomputed set, so the global
rate is largely their own (against non-lanthanide materials the gap is 11.8\% against
14.8\%). And the marginal contrast is almost entirely composition: once anion class is
held fixed the regression gives an odds ratio of 1.02 [0.93, 1.12], with the sign of
the association reversing between anion classes (Section~\ref{sec:ln}). And the
hypothesis was itself informed by the earlier unstratified inspection of these data
(Section~\ref{sec:contrib}), so its confirmation is not an out-of-sample prediction. The
pre-committed claim is confirmed as stated and uninformative as posed.

The rules and the results were committed in that order, at \texttt{82d49db} and
\texttt{c11ccf2} respectively, in the public repository. This is a sole-authored
repository and git commit timestamps are set by the committer, so the ordering is
asserted by the author and attested by nobody; it should be read as a description of how we
worked rather than as independent certification. What the repository does establish is
content: the decision rules in \texttt{DECISION\_RULES.md} are those paraphrased above, and
the thresholds were not adjusted once the spread was known.

\section{Results}

\subsection{The aggregate figure reproduces, and conceals a 26.7-point spread at fixed element count}\label{sec:cells}

Across all 139{,}541 recomputed materials the flip rate is 13.04\% against the Materials
Project hull, 15.93\% against MP+OQMD, and 25.8\% against the most inclusive reference
set. Validating against the source requires matching its populations rather than its
headline. Restricted to quaternaries our survival rate (one minus the flip rate) is 87.35\% against the reported
86.8\%, a gap of about half a point; pooled over binaries and ternaries it is 85.34\% against
the reported 84\%, a gap of 1.3 points. The source reports these shares without sample
sizes or the reference hull behind them, so agreement to within about one point is the
most this comparison can establish; the residual gaps are consistent with the reference
hull having been rebuilt since publication, and we do not claim exact agreement.

\begin{figure}[tbp]\centering
\begin{minipage}[t]{0.52\textwidth}\centering
\includegraphics[width=\textwidth]{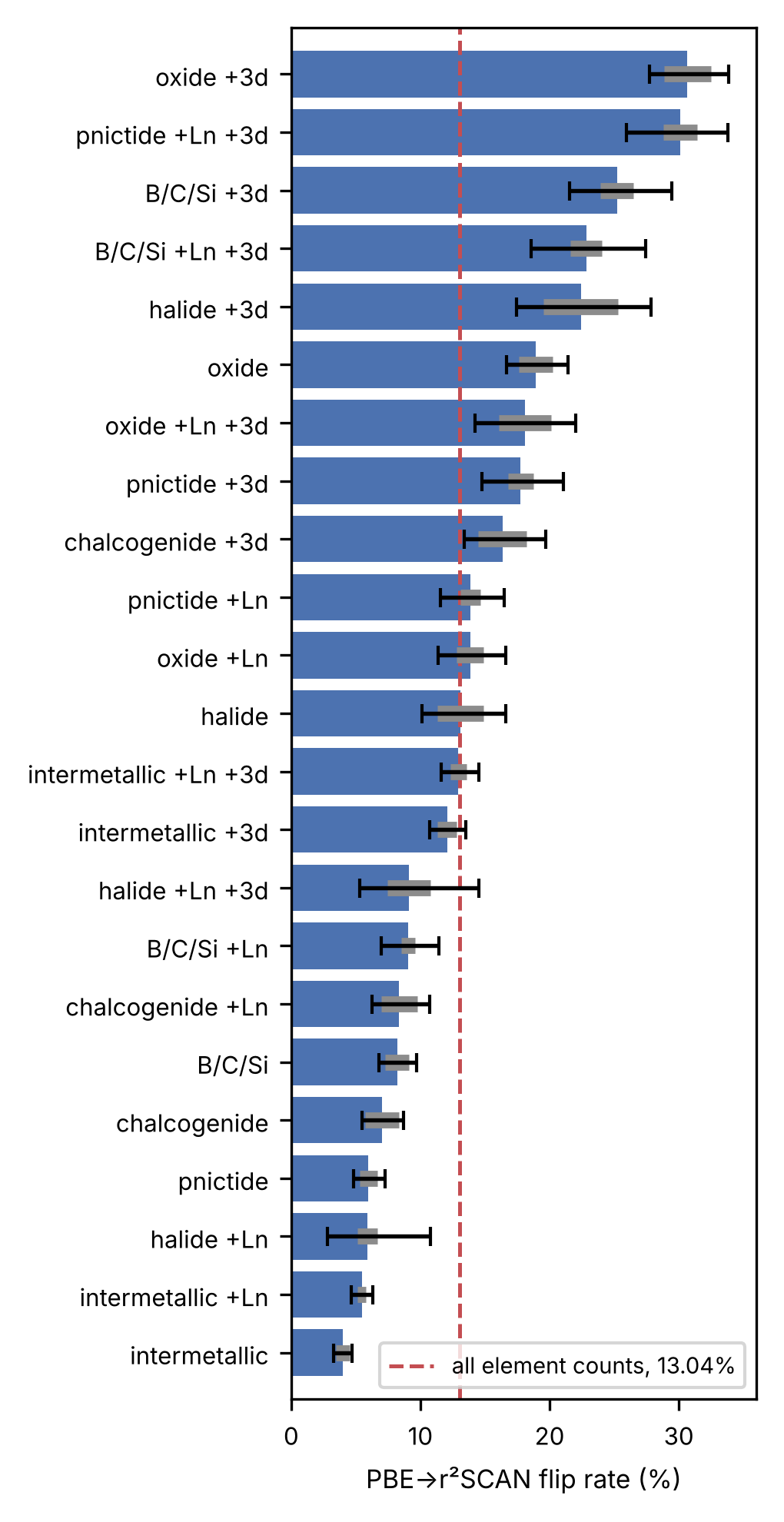}
\end{minipage}\hfill
\begin{minipage}[t]{0.46\textwidth}\centering
\includegraphics[width=\textwidth]{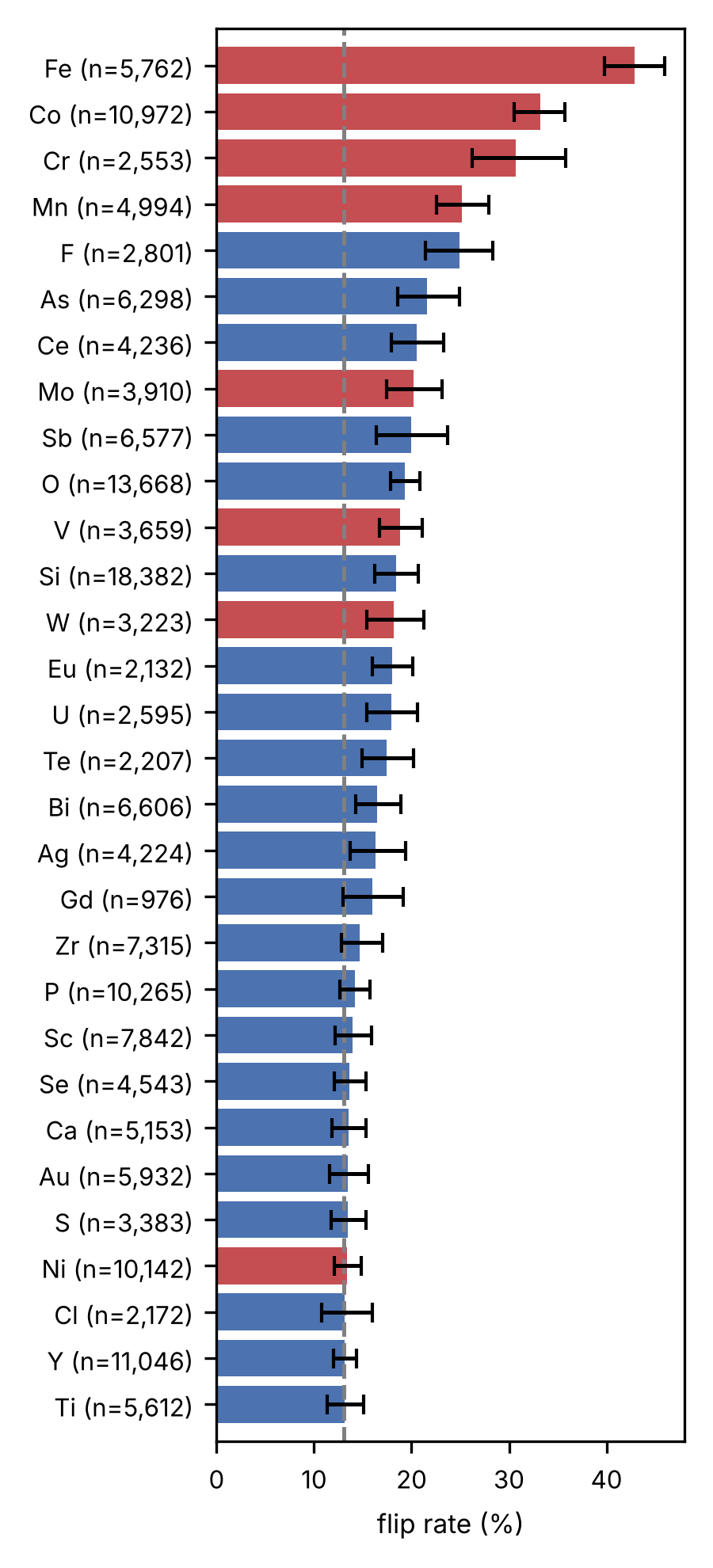}
\end{minipage}
\caption{\textbf{Left:} flip rate by composition cell for quaternary GNoME materials.
Cells are disjoint; B/C/Si abbreviates boride/carbide/silicide. Black bars are cluster bootstrap 95\% intervals; the grey bars behind
them are the naive Wald binomial intervals, shown to make the size of the correction visible.
The dashed line in both panels marks the flip rate over all element counts, 13.04\%,
as labelled in the left-panel legend; the quaternary-only marginal is 12.65\%. \textbf{Right:} per-element flip rate, 30
highest of 83 elements with $n\ge300$, with cluster bootstrap 95\% intervals. Red bars mark
the eight elements receiving a Hubbard $U$ in oxides and fluorides in the MP-compatible workflow.}
\label{fig:chem}
\end{figure}

Resolved by chemistry at fixed element count, the picture changes entirely
(Figure~\ref{fig:chem}, left; full table with cluster bootstrap intervals in
Appendix~\ref{app:cells}). Among quaternary materials, 23 cells qualify and the flip rate
ranges from 4.0\% [3.3, 4.7] for plain intermetallics containing neither a lanthanide nor
a 3$d$ metal ($n=5{,}025$) to 30.7\% [27.7, 33.8] for 3$d$-containing oxides without a
lanthanide ($n=2{,}478$). The spread of 26.7 points is more than five times the
pre-committed threshold and the intervals remain disjoint under clustering. The
highest-minus-lowest spread is a selected extremum and we report it descriptively: the
pre-committed rule fixed the decision threshold before the analysis was run, but it does
not make the realised pair a prespecified pairwise contrast, and we attach no
multiplicity-corrected $p$-value to it. Ternaries show a comparable spread of 23.9 points
across 17 cells. The ordering is compatible with prior chemistry-dependent benchmarking
\citep{kothakonda2023} without being predicted by it, since comparable accuracy against
experiment does not imply agreement between two workflows.

\subsection{Per-element rates against the MP reference span a factor of thirty, 7.5 among non-\texorpdfstring{$f$}{f} elements}\label{sec:element}

Computing the flip rate for every element appearing in at least 300 recomputed materials
(83 elements; Figure~\ref{fig:chem}, right; full table in Appendix~\ref{app:element}) gives a
sharper picture. Iron-containing materials flip at 42.8\% [39.8, 46.0] ($n=5{,}762$,
1{,}218 prototype clusters). Cobalt follows at 33.2\% [30.5, 35.7], chromium at 30.7\%
[26.2, 35.8] and manganese at 25.2\% [22.6, 27.9]. At the other end, the lowest are
actinium at 1.4\% [0.9, 2.0], protactinium at 3.6\% [2.6, 4.8] and promethium at 4.4\% [0.8, 9.3]
($n=389$, an interval that overlaps the next entries), with technetium and magnesium tied at 5.7\%. Of the 83 element-level contrasts
against the global rate, 53 remain significant after Benjamini--Hochberg correction at
the 0.05 level, and at 20{,}000 replicates the same 53 survive under every bootstrap seed
we tried (Appendix~\ref{app:impl}). The 30 that do not are the elements whose
element-minus-global contrast is not distinguishable
from zero; that is a statement about the difference and not about the rate itself.

Three cautions apply. Per-element rates are not disjoint: an iron oxide counts toward
both Fe and O, so an element's rate is confounded with the anion classes in which it
tends to occur, and Section~\ref{sec:u} reports instead the \emph{excess} flip rate
associated with Fe, Co and Ni within each anion class. The two lowest entries are
actinides (Ac, Pa), which carry an $f$-electron caveat analogous to the lanthanides'
(Section~\ref{sec:ln}), so the factor-of-thirty range should not be over-read; among non-$f$
elements it is a factor of 7.5. And the correspondence with the Materials Project $U$
list is imperfect: nickel sits at 13.4\%, essentially the global average, and
Section~\ref{sec:u} shows its excess does not sit where a $U$-suppression account
predicts.

\subsection{Elevated disagreement persists outside the Hubbard-\texorpdfstring{$U$}{U}-treated regime}\label{sec:u}

PBE labels carry the MP2020 corrections; the r$^2$SCAN side carries no documented
correction (Section~\ref{sec:datasets}). If the gradient
simply tracked where those corrections are applied, it would be an artefact of
bookkeeping. The anion corrections \citep{mpanion} attach to the anion rather than to
the transition metal, so comparing Fe-containing against Fe-free materials of the same
anion class differences out most of their effect, though not exactly, since whether a
species receives the correction depends on its chemical role. The Hubbard $U$ is the
opposite: it applies to V, Cr, Mn, Fe, Co, Ni, W and Mo, only in oxides and fluorides
\citep{warford2026,mpanion}, and it acts on the cation, which is why a within-class
Fe-present against Fe-absent contrast is the sharpest handle on it the released data
offer. We recompute no corrections anywhere in this work. Three comparisons examine
whether the association follows the $U$ boundary.

\paragraph{First: the regime with no $U$.} Among compositions containing neither oxygen
nor fluorine, quaternary materials containing one of the eight $U$-list elements flip at
21.9\% ($n=29{,}482$), those whose only 3$d$ elements are Sc, Ti, Cu or Zn at 7.2\%
($n=14{,}649$), and those with no 3$d$ element and no Mo or W at 7.4\% ($n=54{,}795$); ternaries give
28.4\%, 10.8\% and 9.6\% ($n=5{,}245$, 3{,}102 and 15{,}765). The same three groups
within oxides and fluorides, where the $U$ is applied, flip at 32.8\%, 19.7\% and 13.7\%
for quaternaries ($n=4{,}009$, 1{,}246 and 7{,}633). The groups are mutually
exclusive and are the same indicators as the regression's 3$d$ encoding
(Appendix~\ref{app:impl}). The Sc/Ti/Cu/Zn group shows an excess confined to oxides and
fluorides (19.7\% against 13.7\%, but 7.2\% against 7.4\% elsewhere) that neither the
$U$ reading nor the magnetism reading below covers; we list it with
molybdenum and tungsten as unexplained.

\paragraph{Second: across the $U$ boundary directly.} Within each anion class
(Table~\ref{tab:feexcess}), the iron excess is smallest in oxides, where Fe receives a
$U$, and largest in borides, pnictides and intermetallics, where it does not; the oxide
and boride intervals are far from overlapping. Cobalt's oxide excess is likewise among its smallest and indistinguishable from zero
($+4.6$ [$-1.0$, $+11.7$]); its halide excess is smaller still. Nickel runs the other
way: its largest excesses are in chalcogenides ($+14.2$ [$+6.9$, $+23.8$]) and oxides
($+10.3$ [$+3.8$, $+17.5$]), with a small excess in intermetallics ($+2.0$ [$+0.3$,
$+3.7$]) and none distinguishable from zero in pnictides, borides or halides. The oxide iron
excess, $+5.0$ [$+0.5$, $+10.1$], is the iron interval with the least margin from zero; its lower
endpoint is stable across bootstrap seeds (Appendix~\ref{app:impl}) but sits close to
zero, so we rest nothing on its excluding zero. What the section argues from is the gap
between that excess and the $+29$ to $+47$ intervals in borides, pnictides and
intermetallics, where iron receives no $U$.

\begin{table}[htbp]\centering
\caption{Excess flip rate associated with the presence of Fe, Co or Ni, within each broad
composition class. The classes are coarse and their members differ in more than the
transition metal, so these are associations rather than causal element effects.
``Without'' is the flip rate of materials in the same class lacking that element; the
excess is in percentage points (pp), computed from unrounded rates. The oxide row, the
only class in which Fe receives a $U$ throughout, is set in bold; fluorides, a minority of
the halide row, also receive it (third comparison below). Intervals are cluster bootstrap
intervals over structural prototypes, 100{,}000
replicates; one draw per replicate serves all three elements. Classes with fewer than 100
element-containing materials are omitted, which removes the Fe-hydride row only. The final
column gives the same contrast as an odds ratio, since a percentage-point excess is
compressed where the baseline is high; the bottom of the ordering --- oxide lowest,
halide next --- is unchanged on either scale.}
\label{tab:feexcess}
\small\begin{tabular}{lrrrcr}
\toprule
Anion class & $n$ with Fe & With Fe & Without Fe & Excess, pp [95\% CI] & OR \\
\midrule
boride/carbide/silicide & 1{,}268 & 54.0\% & 13.0\% & $+41.0$ [$+35.4$, $+47.0$] & 7.88 \\
pnictide                & 1{,}264 & 50.3\% & 15.0\% & $+35.3$ [$+29.8$, $+40.6$] & 5.75 \\
intermetallic           & 1{,}904 & 42.2\% & \phantom{0}8.0\% & $+34.2$ [$+29.4$, $+39.1$] & 8.38 \\
chalcogenide            & \phantom{0}154 & 41.6\% & 11.5\% & $+30.0$ [$+17.7$, $+44.8$] & 5.46 \\
halide                  & \phantom{0}476 & 24.4\% & 10.9\% & $+13.5$ [$+5.8$, $+22.4$] & 2.64 \\
\textbf{oxide} ($U$ applied) & \phantom{0}610 & 24.1\% & 19.1\% & $\mathbf{+5.0}$ [$+0.5$, $+10.1$] & 1.35 \\
\midrule
\multicolumn{6}{l}{Co: boride $+34.0$ [$+28.0$, $+39.6$], intermetallic $+22.2$ [$+18.5$, $+25.9$], pnictide $+22.0$ [$+17.8$, $+25.6$],} \\
\multicolumn{6}{l}{\phantom{Co: }chalcogenide $+15.8$ [$+9.4$, $+24.2$], hydride $+9.7$ [$-0.8$, $+22.9$], oxide $+4.6$ [$-1.0$, $+11.7$],} \\
\multicolumn{6}{l}{\phantom{Co: }halide $-0.7$ [$-6.0$, $+6.4$]} \\
\multicolumn{6}{l}{Ni: chalcogenide $+14.2$ [$+6.9$, $+23.8$], oxide $+10.3$ [$+3.8$, $+17.5$], hydride $+6.0$ [$-4.1$, $+19.4$],} \\
\multicolumn{6}{l}{\phantom{Ni: }intermetallic $+2.0$ [$+0.3$, $+3.7$], pnictide $-1.5$ [$-5.1$, $+2.2$], boride $-2.2$ [$-5.2$, $+1.0$],} \\
\multicolumn{6}{l}{\phantom{Ni: }halide $-2.1$ [$-6.2$, $+4.7$]} \\
\bottomrule
\end{tabular}
\end{table}

\paragraph{Third: fluorides against the other halides.} The MP-compatible input set applies $U$ in oxides
\emph{and fluorides}, so within the halides, fluorine-containing materials receive it on
the $U$-list metals and the chloride, bromide and iodide materials do not. The iron
excess is $+8.8$ points among fluorine-containing halides against $+16.3$ among the
rest, and the manganese excess $+2.4$ against $+21.5$, the direction of the oxide result
in both cases. The subsets are small ($n=78$ Fe-fluorides, 170 Mn-fluorides), the two
halide subsets differ in halogen chemistry as well as in $U$ treatment, and cobalt
fluorides ($n=30$) are too few to split, so this is
corroboration rather than independent evidence. That
fluorine-containing halides flip at 24.8\% overall against 7.2\% for the rest is a
difference in class baseline; the quantity the $U$ argument concerns is the within-class
excess contributed by a $U$-list metal, $+6.3$ points among fluorides and $+7.8$ among
the other halides, a contrast that is nearly flat; the per-element contrasts above are
what carry the direction.

\paragraph{Reading.} The version of the selective-$U$ hypothesis tested here is that
disagreement produced by the $+U$ treatment would concentrate in the oxide and fluoride
chemistry where $U$ is applied. Elevated disagreement persists outside that regime, and
the iron-associated within-class excess is largest there (the pooled $U$-list excess
itself is not smaller where $U$ is applied: $+19.1$ points in oxides and fluorides
against $+14.5$ outside, from the first comparison), so
selective $U$ application is not a sufficient explanation. What
follows is an association across composition classes, not a controlled intervention ---
oxides, borides, pnictides and intermetallics differ in bonding, oxidation state,
magnetic behaviour, competing phases and reference coverage, not only in whether a $U$
is applied --- so we write that the smaller excesses in oxides are \emph{consistent with}
partial suppression by the corrected treatment, and we do not claim that applying $U$
causes the gradient to shrink. Reference coverage is a further alternative: on the
Materials Project side the r$^2$SCAN hulls of oxides are the least complete of any class
(Section~\ref{sec:allref}), and incompleteness understates flips for iron-containing and
iron-free oxides alike, compressing the excess. Where $U$ is applied, the iron excess shrinks from the $+34$ to $+41$ range of the
classes receiving no $U$ (and $+13.5$ in the halides, where only fluorides receive it) to $+5.0$; cobalt's oxide excess is small, but so is its halide excess in a
class that mostly receives no $U$; nickel runs the other way (Table~\ref{tab:feexcess}); and on
the continuous scale of Section~\ref{sec:additivity} the oxide context is the lowest for
Fe and Cr but the highest for Co. The partial-suppression reading therefore holds for
iron, is ambiguous for cobalt, and does not hold for the $U$ list as a whole. The highest within-class rates in
Table~\ref{tab:feexcess} are iron in borides, pnictides and intermetallics, chemistry that receives no $U$, which is where Warford et al.\ identify the
uncorrected potential-energy surface \citep{warford2026}; this is a statement about $U$
specifically, since chalcogenides receive a full anion correction and still show a
$+30.0$-point iron excess.

\paragraph{What the operative variable might be.} An element-presence indicator cannot
separate Hubbard-$U$-list membership from 3$d$ magnetism. One mechanism is well
documented and is a plausible candidate for what we observe, though we compare no
candidates and rank none. SCAN-family meta-GGAs overestimate magnetic moments and
magnetic stabilisation in itinerant 3$d$ systems. For the elemental ferromagnets,
\citet{ekholm2018} report that SCAN ``performs worse than standard local and semilocal
functionals'' for fcc Ni and hcp Co. \citet{fusingh2018} describe a degradation relative
to PBE across V, Co, Ni and Pd, and in Fe$_3$C obtain 6.87 $\mu_B$ per three-iron formula
unit under SCAN against 5.75 under PBE and a room-temperature saturation value of 5.3, an
excess of about 30\% over experiment. \citet{mejia2019} trace the elemental
over-magnetisation, about 20\% in bcc Fe, to the iso-orbital indicator;
\citet{tran2020} find it in the intermetallic ferromagnets they test; and
\citet{kothakonda2023} find it stronger under r$^2$SCAN than under SCAN, with iron's
moment overestimated by 24\% against SCAN's 17\%. A functional that
over-stabilises the magnetic state displaces a metallic compound and its competing
phases by different amounts, which is what a decomposition energy measures. Three
features of our results are consistent with that: the Fe, Co, Cr, Mn ordering at the top
of the per-element rates; the concentration of the iron and cobalt excess in borides,
pnictides and intermetallics, classes that are predominantly metallic; and its
near-absence in oxides, where moments are typically more localised under both
treatments. These are class-level tendencies, not material-level evidence. Nickel's position near the global average
is also consistent, because its spin-up $d$ bands are virtually filled already under
PBE \citep{ekholm2018}, though that accounts for the average and not for its reversed
anion-class pattern. Molybdenum and tungsten at 20.2\% and 18.2\% are non-magnetic 4$d$
and 5$d$ metals and we leave them unexplained. Because the release initialises every
calculation ferromagnetically \citep{merchant2023}, the two workflows may additionally
converge to different magnetic states. We have not tested this mechanism and do not
claim it: the released files carry no magnetisation field, and the discriminating
calculation is a spin-polarised against non-spin-polarised comparison on a small set of
iron and cobalt borides and pnictides with their competing phases.

Separately, we asked which element set best discriminates the released outcomes, refitting a reduced form of the
regression of Section~\ref{sec:logit} with the MP $U$ list, its 3$d$ subset alone, and
an alternative 3$d$ indicator that drops Mo and W and adds Cu. The Cu-including
indicator fits worst (log-likelihood $-50{,}590$ against $-50{,}472$ for the $U$ list
and $-50{,}329$ for its 3$d$ subset); copper-containing materials flip at 12.6\%, below
the global 13.0\%, so adding copper dilutes the indicator. The 3$d$ subset fits best of
the three at equal parameter count. We nevertheless keep the full $U$ list as the
headline encoding because it names the correction scheme's own element set and is the
encoding used throughout; the difference between the two is a
matter of Mo and W. We did not resolve oxidation
states, so this concerns which element set best discriminates the outcomes and not electronic
occupancy.

\subsection{Element coefficients vary with anion context in the fitted model}\label{sec:additivity}

A natural objection is that the chemistry dependence might be per-element
bookkeeping: if r$^2$SCAN's elemental reference energies were offset for the magnetic
3$d$ metals, every compound containing them would shift. That mechanism is excluded by
construction, because a decomposition energy is evaluated at fixed composition: an
additive shift $\delta_e$ in element $e$'s reference contributes $\sum_e x_e \delta_e$,
with $x_e$ the atomic fraction of element $e$, to
the compound and the identical amount to the competing combination, and cancels. What the
argument does not cover is an error in the computed energy of a particular competing
phase, including an elemental metal that is itself a decomposition product: an
over-stabilised bcc iron raises the decomposition energy of every compound that decomposes
to it, and does not cancel. The broader question is whether $\Delta$, the per-material r$^2$SCAN-minus-PBE
displacement of the decomposition energy defined in Section~\ref{sec:shift}, is an
additive function of composition.
A naive test fails here: our cells are \emph{defined} by element presence, so an additive
per-element model spans nearly the same space as the cell indicators, and removing its fit
collapses the cell gradient in $\Delta$ essentially to zero --- collinearity, not
evidence.
The discriminating question is whether a given element contributes differently
in different company. Allowing Fe, Co, Ni, Mn and Cr to interact with anion class
improves the fit by $\Delta R^2 = 0.009$ on $n=117{,}018$ with 30 added terms, and the
fitted contributions spread widely (Table~\ref{tab:context}): iron's is $-31$ meV/atom per
unit atomic fraction in oxides and $+208$ in chalcogenides, a spread of 239 meV/atom
where the additive model constrains it to be constant; at a realistic iron fraction of
0.1 to 0.25 that is 24 to 60 meV/atom between contexts, between one-and-a-half and four times the cell-level
mean shifts of Section~\ref{sec:shift}. The design is full rank (113 columns, condition
number $7.0\times10^{2}$), so every within-element contrast is identified
(Appendix~\ref{app:impl}). We read the table as a descriptive fit rather than a
calibrated test of additivity, because $\Delta$ contains the correction removal by
construction, so a fitted level is not an element's physical contribution.

\begin{table}[htbp]\centering
\caption{Fitted contribution to $\Delta$ per unit atomic fraction (meV/atom), by element
and anion context, from the ordinary-least-squares model of Appendix~\ref{app:impl}:
$\Delta \sim \sum_e c_e x_e + \sum_{e \in S}\sum_{a} \gamma_{ea} x_e
\mathbb{1}[\text{class}=a]$, with $x_e$ the atomic fraction of element $e$, $S = \{$Fe,
Co, Ni, Mn, Cr$\}$ and $a$ the six non-hydride anion classes. The entry shown is
$c_e + \gamma_{ea}$, and an additive model constrains each row's spread to be zero in the
population. The spread interval is a cluster bootstrap over structural prototypes
(Appendix~\ref{app:impl}); between the two seeds the endpoints move by up to 13 meV/atom,
so they should be read to about $\pm 10$. The interval describes
uncertainty in a fitted spread and is not a test of additivity; for that, see
Table~\ref{tab:jointtest}.}
\label{tab:context}
\small\begin{tabular}{lrrrrrrrr}
\toprule
Element & oxide & halide & chalc. & pnictide & bor./carb./sil. & interm. & Spread & 95\% CI \\
\midrule
Fe & $-31$ & $+195$ & $+208$ & $+143$ & $+157$ & $+80$ & 239 & [161, 334] \\
Cr & $+8$ & $+58$ & $+257$ & $+108$ & $+119$ & $+178$ & 249 & [144, 404] \\
Mn & $+62$ & $+263$ & $+196$ & $-10$ & $+45$ & $+40$ & 273 & [196, 338] \\
Ni & $+84$ & $+71$ & $+27$ & $-8$ & $-3$ & $+19$ & \phantom{0}92 & \phantom{0}[45, 206] \\
Co & $+92$ & $+86$ & $+78$ & $+56$ & $+70$ & $+67$ & \phantom{0}36 & \phantom{00}[23, 97] \\
\bottomrule
\end{tabular}
\end{table}

For Fe and Cr the oxide context is the lowest of all, the same context-dependent
association reported in Section~\ref{sec:u} expressed on the continuous scale and not
independent confirmation of it; for Co the oxide context is the highest ($+92$), so the
suppression pattern is not reproduced for cobalt on this scale; the two elements with the smallest fitted spreads are Co
and Ni, at 36 [23, 97] and 92 [45, 206] meV/atom against 239 to 273 for Fe, Cr and Mn.

Because each spread is a data-selected maximum-minus-minimum statistic, the bootstrap
intervals are not calibrated tests of the additive null. For that we use a
cluster-robust Wald test that an element's six interaction coefficients are equal, five
degrees of freedom each, with Benjamini--Hochberg correction across the five elements
(Table~\ref{tab:jointtest}). Within this post-selected element set and fitted
specification, four reject equal context coefficients: nickel at $\chi^2 = 94.8$,
manganese at 71.6, iron at 70.3 and chromium at 31.7. Cobalt does not, at $\chi^2 = 8.2$
and $p = 0.15$, even though its spread interval excludes zero --- exactly the
over-rejection a selected-range statistic produces.

\begin{table}[htbp]\centering
\caption{Cluster-robust Wald test of the joint null that an element's six interaction
coefficients are equal, against the alternative that at least one anion context differs.
Five degrees of freedom each; the covariance is the cluster-robust sandwich over the
7{,}822 structural prototypes with the finite-cluster correction of
Appendix~\ref{app:impl}, which is immaterial at this cluster count and omitted in the
logistic model; the threshold is
Benjamini--Hochberg (BH) at 0.05 across the five elements, applied in ascending $p$
order.}
\label{tab:jointtest}
\small\begin{tabular}{lrrr}
\toprule
Element & $\chi^2_5$ & $p$ & BH threshold \\
\midrule
Ni & 94.75 & $6.8\times10^{-19}$ & 0.010 \\
Mn & 71.63 & $4.7\times10^{-14}$ & 0.020 \\
Fe & 70.31 & $8.8\times10^{-14}$ & 0.030 \\
Cr & 31.71 & $6.8\times10^{-6}$  & 0.040 \\
Co & \phantom{0}8.15 & 0.148 & 0.050 \\
\bottomrule
\end{tabular}
\end{table}

\subsection{The gradient survives within every populated hull-distance band}\label{sec:hulldist}

The most obvious alternative explanation for Section~\ref{sec:cells} is proximity.
Materials close to the PBE hull need only a small energy change to cross it; if 3$d$
oxides sit nearer the hull than intermetallics, they will flip more often for no reason
connected to the functional, and Supplementary Note 2 of the source work reports that
disagreement concentrates near the hull. Testing it was pre-committed as one of the robustness gates listed in
\texttt{DECISION\_RULES.md}.

Table~\ref{tab:hulldist} stratifies quaternary materials by PBE decomposition energy.
Flip rate rises steeply toward the hull in every cell (monotonically in all but the
3$d$-free intermetallic column, whose two deepest bands are 0.8\% and 0.5\%), confirming the
Note 2 observation: 3$d$-free intermetallics (lanthanide pooled) go from 0.5\% in the $(-0.1,-0.05]$ band to 15.9\% in
$(-0.01,0]$. And the chemistry gradient is present inside every band: in
$(-0.05,-0.02]$, 3$d$ oxides flip at 23.1\% and 3$d$-free intermetallics at 1.2\%; in
$(-0.01,0]$, 40.1\% against 15.9\%. The within-band spread between those two cells is
roughly 11, 22, 25 and 24 points across the four populated bands. The cells do differ in position (median PBE decomposition energy $-20.8$ meV/atom for
3$d$ oxides against $-22.4$ for 3$d$-free intermetallics).

\begin{table}[htbp]\centering
\caption{Quaternary flip rate by PBE decomposition energy band and composition cell.
Cells with fewer than 200 materials in a band are shown as a dash; counts in
parentheses. These
cells pool the lanthanide flag, because the lanthanide-resolved cells are too thin to
stratify further by band, so the oxide+3$d$ minus 3$d$-free-intermetallic contrast here is
smaller than the 26.7 points of Section~\ref{sec:cells}. Column counts differ from the quaternary cell totals of Section~\ref{sec:ipw} by at
most ten materials per cell.}
\label{tab:hulldist}
\footnotesize\setlength{\tabcolsep}{3pt}
\begin{tabular}{lrrrrrr}
\toprule
PBE band (eV/atom) & oxide+3$d$ & oxide & pnictide+3$d$ & pnictide & interm.+3$d$ & interm. \\
\midrule
$\le-0.1$ & --- & 3.9\% (563) & 5.5\% (494) & 0.5\% (794) & 2.0\% (402) & 0.8\% (930) \\
$(-0.1,-0.05]$ & 11.1\% (443) & 6.1\% (1{,}063) & 14.6\% (1{,}748) & 1.3\% (2{,}364) & 3.2\% (1{,}987) & 0.5\% (2{,}710) \\
$(-0.05,-0.02]$ & 23.1\% (1{,}070) & 15.5\% (2{,}132) & 19.8\% (3{,}608) & 8.2\% (3{,}810) & 7.9\% (6{,}709) & 1.2\% (7{,}777) \\
$(-0.02,-0.01]$ & 29.4\% (744) & 24.3\% (1{,}159) & 28.3\% (1{,}752) & 20.8\% (1{,}970) & 16.0\% (3{,}947) & 4.3\% (4{,}512) \\
$(-0.01,0]$ & 40.1\% (802) & 27.8\% (1{,}142) & 37.7\% (1{,}493) & 27.0\% (1{,}442) & 27.2\% (3{,}263) & 15.9\% (4{,}937) \\
\bottomrule
\end{tabular}
\end{table}

Binning leaves residual variation in distance inside each band, so Table~\ref{tab:hullcont}
removes the bins. It refits the composition model of Section~\ref{sec:logit} with a
natural cubic spline in the continuous PBE decomposition energy, on the 117{,}018
materials for which that quantity is observed and non-positive, with cluster-robust
standard errors over the 7{,}822 structural prototypes present in that subset. The
covariate is the summary file's MP-referenced decomposition energy, so both sides of the
comparison use the same reference set (the two hulls themselves differ by functional). The adjustment is substantial and the chemistry
association survives it. Hull distance more than doubles the model's pseudo-$R^2$, from
0.068 to 0.156, so proximity carries a great deal of information that composition does
not; the $U$-list odds ratio nevertheless falls only from 3.53 [3.19, 3.92] to 3.43
[3.07, 3.83]. Every anion-class term moves the other way, oxide from 2.63 to 3.16 and pnictide from
1.93 to 2.58, consistent with those classes, taken as a whole, sitting farther from the hull than the reference
class, so that the unadjusted rate understates their association and
adjusting for position strengthens rather than explains it. Adding
interactions between hull proximity and chemistry shows the two interact:
$U$-list by proximity is 0.52 [0.43, 0.63] and oxide by proximity 0.62 [0.50, 0.76], where
proximity increases toward the hull, so the chemistry multiplier compresses as the hull is approached while
the gap in percentage points widens up to the penultimate band and is roughly flat in
the last, because the baseline rate rises much faster.

\begin{table}[htbp]\centering
\caption{Composition odds ratios with and without a flexible continuous adjustment for
distance below the PBE hull, on the 117{,}018 materials with an observed non-positive PBE
decomposition energy. The spline is a five-column natural cubic basis (truncated-power form with linear
tails) with knots at the sextiles of that energy, columns standardised. Intervals are cluster-robust over the 7{,}822
structural prototypes in this subset. The
reference class is a lanthanide- and 3$d$-free hydride or intermetallic; element-count
and log-sites terms are included in both fits but not shown.}
\label{tab:hullcont}
\small\begin{tabular}{lrr}
\toprule
 & no hull-distance term & natural spline in PBE hull distance \\
\midrule
$U$-list element present   & 3.53 [3.19, 3.92] & 3.43 [3.07, 3.83] \\
other 3$d$ present         & 1.12 [0.99, 1.25] & 1.07 [0.96, 1.20] \\
lanthanide present         & 1.12 [1.01, 1.23] & 1.01 [0.91, 1.12] \\
oxide                      & 2.63 [2.27, 3.05] & 3.16 [2.69, 3.72] \\
pnictide                   & 1.93 [1.69, 2.20] & 2.58 [2.22, 2.98] \\
halide                     & 1.27 [0.98, 1.65] & 1.97 [1.51, 2.59] \\
chalcogenide               & 1.40 [1.18, 1.65] & 1.66 [1.38, 2.00] \\
boride/carbide/silicide    & 1.78 [1.51, 2.09] & 2.34 [1.94, 2.83] \\
\midrule
McFadden pseudo-$R^2$      & 0.068 & 0.156 \\
\bottomrule
\end{tabular}
\end{table}

\subsection{\texorpdfstring{r$^2$SCAN}{r2SCAN} shifts chemistries in opposite directions}\label{sec:shift}

A flip is a binary summary of a continuous quantity,
$\Delta = E_{\text{dec}}(\text{r}^2\text{SCAN}) - E_{\text{dec}}(\text{PBE})$, where the
subtrahend carries the MP2020 corrections and the Hubbard-$U$ treatment and the minuend
carries no documented correction. $\Delta$ is therefore the net displacement from corrected GGA(+$U$) to uncorrected
r$^2$SCAN, not a pure functional difference, and we write ``r$^2$SCAN moves X'' as
shorthand for that net displacement. The distributions differ between chemistries in
both mean and dispersion (Table~\ref{tab:shift}). The mean changes sign: r$^2$SCAN
destabilises 3$d$ oxides by $+14.7$ meV/atom on average and stabilises plain
intermetallics by $-8.9$, with every 3$d$-containing cell sitting above its 3$d$-free
counterpart, and the share of materials moving upward broadly tracks the mean, from 61.1\% to
31.0\%. The within-cell standard deviations, 23.8 to 55.7 meV/atom, exceed the magnitudes of
the mean displacements in every displayed cell, and the quartiles straddle zero in every
cell: this is
a displacement of a wide distribution, not a uniform translation, and for any individual
material $\Delta$ is dominated by scatter. Flips occur where an upward shift carries a
near-hull material across zero, which is why they concentrate in the cells with positive
mean $\Delta$. Kothakonda et al.\ find GGA and meta-GGA ``comparable'' for intermetallics
\citep{kothakonda2023}, and Bartel et al.\ find PBE and SCAN similarly accurate on
decomposition reactions generally \citep{bartel2019}; our data add that
the mean displacement differs in sign between chemistries. Whether r$^2$SCAN's
stabilisation of intermetallics or its destabilisation of 3$d$ oxides is the more
accurate direction is a question about experiment that this paper does not address.

\begin{table}[htbp]\centering
\caption{Change in decomposition energy from PBE to r$^2$SCAN, on the primary cell
definition (quaternary, no lanthanide), restricted to the 117{,}018 materials that carry a
PBE decomposition energy and are PBE-stable. Cell definitions are identical to
Section~\ref{sec:cells}; counts are correspondingly 6\% to 20\% smaller. Positive values
mean r$^2$SCAN moves the material toward instability. All entries are in meV/atom; Q1
and Q3 are the quartiles of $\Delta$.}
\label{tab:shift}
\small\begin{tabular}{lrrrrrrr}
\toprule
Cell & $n$ & mean & median & s.d. & Q1 & Q3 & share $\Delta>0$ \\
\midrule
oxide + 3$d$          & 2{,}085 & $+14.7$ & $+5.4$ & 55.7 & $-8.3$ & $+28.6$ & 61.1\% \\
bor./carb./sil. + 3$d$ & 4{,}167 & $+12.9$ & $+6.4$ & 35.6 & $-5.9$ & $+26.8$ & 60.6\% \\
chalcogenide + 3$d$   & 1{,}272 & $+8.1$  & $+2.6$ & 37.9 & $-6.6$ & $+14.5$ & 56.8\% \\
oxide                 & 2{,}748 & $+6.8$  & $+3.1$ & 49.5 & $-11.1$ & $+22.2$ & 57.1\% \\
pnictide + 3$d$       & 5{,}145 & $+4.5$  & $+0.7$ & 39.1 & $-10.7$ & $+17.5$ & 51.5\% \\
halide + 3$d$         & \phantom{0}658 & $+3.4$ & $-0.8$ & 36.0 & $-14.1$ & $+15.9$ & 49.1\% \\
intermetallic + 3$d$  & 7{,}023 & $-0.1$  & $-1.5$ & 28.7 & $-10.1$ & $+7.4$ & 44.5\% \\
halide                & 1{,}126 & $-3.8$  & $-4.8$ & 29.8 & $-16.2$ & $+6.6$ & 38.2\% \\
bor./carb./sil.        & 3{,}120 & $-3.8$  & $-1.9$ & 31.4 & $-13.7$ & $+11.6$ & 44.0\% \\
chalcogenide          & 1{,}234 & $-4.6$  & $-2.7$ & 27.7 & $-14.2$ & $+5.5$ & 39.5\% \\
pnictide              & 4{,}043 & $-6.2$  & $-2.7$ & 31.2 & $-14.1$ & $+5.9$ & 40.8\% \\
intermetallic         & 4{,}628 & $-8.9$  & $-4.1$ & 23.8 & $-14.2$ & $+1.5$ & 31.0\% \\
\bottomrule
\end{tabular}
\end{table}

\subsection{The 3\texorpdfstring{$d$}{d} contrast persists across band-gap bins}\label{sec:gap}

Band gap is a natural candidate for mediating the effect, since $U$-list compounds are
disproportionately narrow-gap. The gap is recorded in the summary file rather than the
r$^2$SCAN file, so it exists only for materials present in both. Of the 139{,}541
recomputed materials, 22{,}498 have no summary-file row and a further 2{,}320 have the
field empty, neither of which we treat as insulators; 2{,}177 carry an infinite sentinel;
112 are stored as the literal \texttt{0.0}, which the release does not distinguish from
a value rounded or clipped to zero and which form a bin too small to split on 3$d$
content; and 112{,}434 carry a finite positive gap and enter Table~\ref{tab:gap}. A
single-threshold split loses most of the variable's structure, since the marginal rate is
not monotone in the cut point (entries with gaps below 0.01 eV flip at 13.6\% and
those between 0.01 and 0.5 eV at 9.6\%, two groups the first bin of Table~\ref{tab:gap}
merges), so
we resolve it into bins.

\begin{table}[htbp]\centering
\caption{Flip rate by PBE band-gap bin, split on 3$d$ content. Descriptive; no intervals
are attached and no claim of equality across bins is made.}
\label{tab:gap}
\small\begin{tabular}{lrrrrr}
\toprule
Band gap (eV) & $n$, no 3$d$ & Flip & $n$, 3$d$ & Flip & 3$d$ excess (pp) \\
\midrule
0--0.5 & 58{,}958 & \phantom{0}8.5\% & 41{,}097 & 19.0\% & $+10.5$ \\
0.5--1 & \phantom{0}1{,}834 & \phantom{0}8.7\% & \phantom{0,}901 & 18.3\% & \phantom{0}$+9.6$ \\
1--3   & \phantom{0}4{,}719 & 15.0\% & \phantom{0}2{,}166 & 25.3\% & $+10.4$ \\
$>3$   & \phantom{0}2{,}208 & 18.9\% & \phantom{0,}551 & 30.7\% & $+11.8$ \\
\bottomrule
\end{tabular}
\end{table}

Reading across the last column, the 3$d$ excess lies between $+9.6$ and $+11.8$ points
in every displayed bin, so the 3$d$-associated contrast persists across the bins.
Reading down the no-3$d$ column, the rate more than doubles from 8.5\% at narrow gaps to
18.9\% above 3 eV, so a gap association is present at fixed 3$d$ content. These are four
descriptive bin comparisons with no intervals attached; the 3$d$ wide-gap cell is the least
well populated in this table ($n=551$), and other
composition differences remain within bins --- among the wide-gap, 3$d$-free materials the
halides flip at 27.4\% and the oxides at 16.5\%, so anion class may carry part of the
gap trend. We report band gap as a second axis along which
disagreement varies and do not attempt to adjudicate between the two.

\subsection{Selection into the recomputed subset}\label{sec:ipw}

The recomputed subset is not a random sample. We fit a logistic model of recomputation
on composition, element count, cell size and PBE decomposition energy (linear and
quadratic) across the 371{,}870 $\le$4-element materials, of which 116{,}814 (31.4\%)
were recomputed, 116{,}789 after removing the 25 PBE-unstable entries. Selection is chemistry-dependent: oxides are more likely to be
recomputed (odds ratio 1.64), as are borides, carbides and silicides (1.58) and
pnictides (1.43), while lanthanide-containing materials (0.68) and larger cells (0.65
per standard deviation of $\log$ sites) are less likely.

The contrast changes little under the specified propensity model. The naive rate over the
116{,}789 recomputed, PBE-stable $\le$4-element entries is 13.36\%, and the weighted
estimate on that IPW-eligible population is 12.71\%. The spread across lanthanide-pooled quaternary cells moves
from 22.0 to 21.9 percentage points (Table~\ref{tab:ipw}); cell-level estimates move by
at most 2.4 points, the ordering of the seven tabulated cells is preserved, and two
untabulated cells each swap with an adjacent tabulated cell. Inverse-probability weighting corrects
only for selection on the observed features entered into the model; if the authors
selected on something we cannot see, it does not correct for it. The weights need their
own accounting (Appendix~\ref{app:impl}): fitted propensities over the 365{,}504 PBE-stable $\le$4-element entries fall to the
underflow limit of the fit, $e^{-500} \approx 7.1\times10^{-218}$ (on entries that were not recomputed; among recomputed entries the
minimum is $1.52\times10^{-6}$), indicating severe estimated lack of overlap, and the
estimate depends on a propensity floor of $10^{-2}$, below which
26 recomputed entries fall. Untrimmed, Kish's effective sample size collapses to 5; at the floor it is
83{,}875 (71.8\% of $n$), and floors of $2\times10^{-2}$ and $5\times10^{-2}$ give
12.74\% and 12.76\%. We report 12.7\% as a sensitivity result conditional on this model
and this floor, not as a corrected population rate.

\begin{table}[htbp]\centering
\caption{Quaternary cells with the lanthanide flag pooled (as in
Table~\ref{tab:hulldist}), naive against inverse-probability-weighted flip rate. Selected
rows; full table in the repository. Rates are computed on all 116{,}814 recomputed entries
with at most four elements, including the 25 that are not PBE-stable. Spreads are computed from unrounded rates.}
\label{tab:ipw}
\small\begin{tabular}{lrrr}
\toprule
Cell & $n$ & Naive & IPW \\
\midrule
oxide, 3$d$                    & \phantom{0}3{,}178 & 27.2\% & 27.2\% \\
boride/carbide/silicide, 3$d$  & \phantom{0}7{,}947 & 25.2\% & 25.8\% \\
pnictide, 3$d$                 & \phantom{0}9{,}097 & 22.6\% & 24.6\% \\
oxide, no 3$d$                 & \phantom{0}6{,}059 & 16.8\% & 16.7\% \\
intermetallic, 3$d$            & 16{,}318 & 13.0\% & 13.1\% \\
halide, no 3$d$                & \phantom{0}4{,}302 & \phantom{0}7.2\% & \phantom{0}8.1\% \\
intermetallic, no 3$d$         & 20{,}871 & \phantom{0}5.3\% & \phantom{0}5.3\% \\
\midrule
\textbf{Spread} & & \textbf{22.0 pp} & \textbf{21.9 pp} \\
\bottomrule
\end{tabular}
\end{table}

\paragraph{Crystal system and cell size.} By crystal system, flip rates run from 16.8\%
(triclinic) to 11.1\% (hexagonal), with cubic at 14.3\% and the rest between 12.6\% and
13.2\%; the triclinic excess is plausibly a proxy for low-symmetry structures being
harder to relax consistently, but we have not established that. By cell size, rates range
from 11.3\% ($\le$10 sites) to 16.3\% ($>$32 sites), not monotonically; in the regression below the size
term is not significant once composition is controlled for.

\subsection{A logistic model over composition}\label{sec:logit}

Table~\ref{tab:logit} reports a logistic regression of flip probability on composition
features across all 139{,}541 materials.

\begin{table}[htbp]\centering
\caption{Logistic regression of flip probability. Reference category pools intermetallics
with hydrides, both lacking a 3$d$ metal and a lanthanide; hydrides are 1.3\% of the set,
and their flip rate (12.0\%) sits above the 3$d$- and lanthanide-free intermetallic rate
of 4 to 8\% for ternaries and quaternaries (14\% for the 92 binaries). Refitting this specification with the hydride rows dropped moves the $U$-list
odds ratio from 3.448 to 3.468, the other-3$d$ term from 1.101 to 1.104 and the
lanthanide term from 1.023 to 1.022; no odds ratio moves by more than 0.023.
$n=139{,}541$; McFadden pseudo-$R^2 = 0.066$. The exponentiated intercept is the
baseline odds for the reference category, not an odds ratio. Intervals are
cluster-robust (sandwich estimator on structural prototypes). The final column gives the
inflation of the standard error relative to the naive analytic value.}
\label{tab:logit}
\small\begin{tabular}{lrrrr}
\toprule
Feature & Coefficient & Odds ratio$^{*}$ & Cluster-robust 95\% CI & SE inflation \\
\midrule
Intercept (baseline odds)    & $-2.780$ & 0.062 & [0.056, 0.068] & 2.24$\times$ \\
MP $U$-list element          & $+1.238$ & 3.448 & [3.129, 3.799] & 2.77$\times$ \\
Other 3$d$ (Sc, Ti, Cu, Zn)  & $+0.096$ & 1.101 & [0.984, 1.232] & 1.98$\times$ \\
Lanthanide present           & $+0.023$ & 1.023 & [0.933, 1.121] & 2.73$\times$ \\
Oxide                        & $+0.903$ & 2.467 & [2.161, 2.815] & 2.43$\times$ \\
Pnictide                     & $+0.666$ & 1.946 & [1.700, 2.228] & 3.00$\times$ \\
Boride/carbide/silicide      & $+0.518$ & 1.678 & [1.437, 1.959] & 3.32$\times$ \\
Chalcogenide                 & $+0.335$ & 1.399 & [1.205, 1.624] & 1.97$\times$ \\
Halide                       & $+0.286$ & 1.330 & [1.047, 1.691] & 3.42$\times$ \\
Element count ($z$-scored)   & $-0.177$ & 0.838 & [0.807, 0.870] & 2.32$\times$ \\
$\log$ sites ($z$-scored)    & $+0.015$ & 1.015 & [0.958, 1.075] & 3.44$\times$ \\
\bottomrule
\multicolumn{5}{l}{\footnotesize $^{*}$ For the intercept row this column is the baseline
odds, not a ratio.} \\
\end{tabular}
\end{table}

The presence of a $U$-list element is associated with 3.45 times the \emph{odds} of a flip. An odds
ratio is not a risk ratio: at the reference baseline of about 6\% the corresponding
probability ratio is 3.0. As an average marginal effect, computed from the interaction
model of the next paragraph with 1{,}000 cluster bootstrap refits, the presence of a
$U$-list element is associated with a flip probability 15.4 points higher [$+14.2$,
$+16.6$], a comparison of $U$-list present against all $U$-list-absent materials pooled.

Lanthanide content has an odds ratio of 1.02 with an interval straddling unity, and that
should not be read as ``lanthanides do not matter''. The model above has no
interactions, so a single lanthanide coefficient averages effects of opposite sign.
Adding lanthanide $\times$ anion and $U$-list $\times$ anion interactions improves the
fit by a likelihood-ratio $\chi^2$ of 879 on 14 degrees of freedom (in a reduced
specification without the other-3$d$ and site-count terms, with hydride as its own class), read as descriptive
since it assumes independent observations. The model-based
average marginal effect of lanthanide content, from the interaction model with
cluster bootstrap intervals, is $-8.3$ [$-11.4$, $-4.8$] in halides, $-7.8$ [$-10.7$,
$-4.9$] in oxides and $+8.9$ [$+6.5$, $+11.6$] in pnictides, while the pooled effect
over all classes is $+0.2$ points [$-0.7$, $+1.2$]; the unadjusted within-class
differences, pooled over all element counts, are $-9.9$, $-9.2$ and $+7.0$ points. The sign reversal is the one reported
nonparametrically in Section~\ref{sec:ln}. Element count enters negatively --- each
standard deviation reduces the odds by about 16\% --- consistent with the marginal rates
(binaries 19.8\%, ternaries 14.6\%, quaternaries 12.7\%).

Three cautions. The McFadden pseudo-$R^2$ is 0.066, low on its own scale: what supports
using composition for allocation is the held-out ranking performance of
Section~\ref{sec:budget}, not this number, and readers should not use this table to
decide whether a specific compound's label is wrong. Correcting for clustering inflates
every standard error by roughly 2 to 3.5 times; the $U$-list association survives with
a wide margin, while the halide term survives only narrowly (1.330 [1.047, 1.691]).
And the $U$-list indicator aggregates eight chemically distinct elements, two of which
(Mo, W) are not 3$d$ at all, so we describe $U$-list membership as a strong observed
compositional predictor and not as an identified physical mechanism.

\subsection{The lanthanide effect reverses sign with the anion}\label{sec:ln}

Lanthanide treatments raise a well-known \emph{accuracy} concern, but that concern does
not by itself predict a high rate of \emph{disagreement} between two workflows that may
share similar approximations, and we keep the two apart. The concern is shared by An
et al., who excluded 142 rare-earth compounds from their stability screen on exactly these
grounds \citep{an2026}, and frozen-$f$ pseudopotentials are documented, for the
norm-conserving PseudoDojo tables, as appropriate only where the $f$ electrons are
chemically unimportant \citep{pseudodojo2018}; that documentation concerns a different
potential family from the PAW datasets VASP uses, and the release does not state its
lanthanide PAW datasets, so the applicability of the concern to this dataset specifically
remains an assumption rather than a checked fact.

Lanthanide-containing materials have a below-global release-based disagreement rate,
11.8\% against 13.0\%, which is the direction we pre-committed on the strength of an
unstratified cut made before the decision rules were written. That does not establish
the adequacy of either workflow: agreement between two workflows cannot rule out an error
they share. And the matched-cell comparison shows the average is misleading. Among
quaternary 3$d$-containing materials, lanthanide-containing cells have \emph{lower}
rates in halides (9.1\% against 22.4\%) and oxides (18.1\% against 30.7\%) and a
\emph{higher} rate in pnictides (30.1\% against 17.8\%); among quaternaries without a
3$d$ metal, the lanthanide-containing pnictide cell sits at 13.9\% against 6.0\%. The
sign depends on the anion. A flip in lanthanide chemistry is not evidence that r$^2$SCAN
is correct, and the absence of one is not evidence that either treatment is adequate:
the disagreement is a flag, not a verdict.

\subsection{Matched against known materials, the aggregate comparison inverts}\label{sec:known}

GNoME materials flip at 13.0\%; Materials Project entries computed at both levels flip
at 14.9\%, so at face value the GNoME entries appear \emph{more} robust to the choice of functional. This
runs opposite to the source paper's own aggregate comparison (84\% survival against the
Materials Project's 90\%), which uses different populations and reference hulls.

\begin{figure}[htbp]\centering
\includegraphics[width=0.71\textwidth]{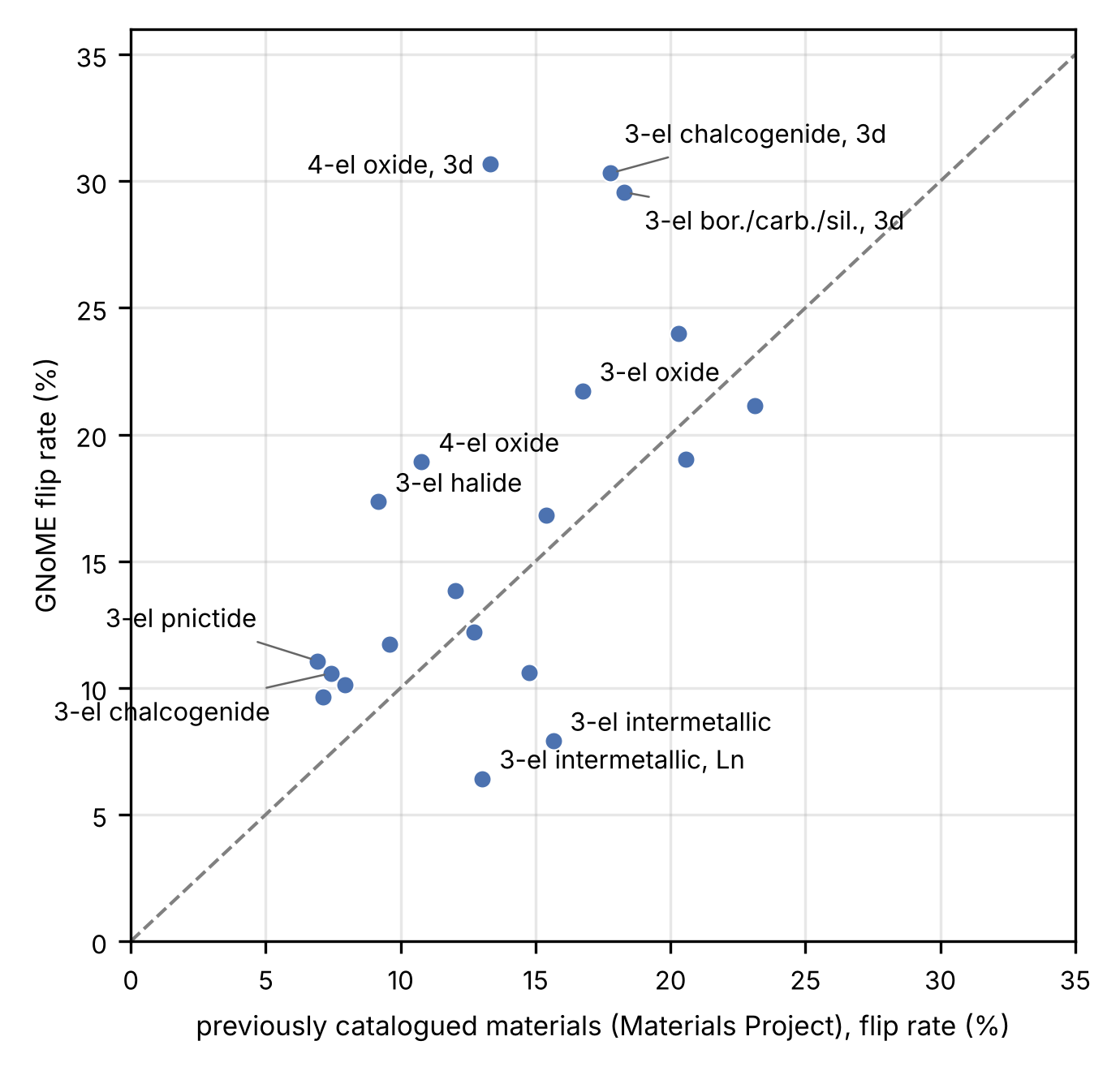}
\caption{Workflow disagreement in matched composition cells. Points above the diagonal
are cells where GNoME materials disagree more than previously catalogued materials of
the same broad composition class and element count; labels mark the ten cells whose
intervals exclude zero, and an unqualified label denotes a lanthanide- and 3$d$-free
cell. Higher disagreement is not lower accuracy.}
\label{fig:vsknown}
\end{figure}

Twenty cells hold at least 500 GNoME and 100 known materials and so enter the
comparison. GNoME disagrees more in 14 and less in 6, which on a two-sided exact sign test is not
itself significant ($p=0.115$); what carries the result is the standardised contrast reported below, together with
the fact that 10 of the 20 differences have intervals excluding zero and show a
chemistry-structured descriptive pattern:
the two negative contrasts whose intervals exclude zero are intermetallic cells
(Table~\ref{tab:known}; all twenty in Figure~\ref{fig:vsknown} and
Appendix~\ref{app:allcells}). The 20 comparisons carry no multiplicity correction, so we
label the comparison exploratory; it is the weakest inferential step in the paper. The
GNoME side is a cluster bootstrap drawn across all twenty cells at once, since 45\% of
the prototypes spanning these cells appear in more than one; the Materials Project side
is resampled at the material level, because the thermo fields retrieved carry no
structural information. Material-level resampling on the GNoME side instead would narrow
the intervals by a median factor of 1.28 and move one cell's interval from including
to excluding zero. The largest deficits are quaternary 3$d$
oxides ($+17.3$ points), ternary 3$d$ chalcogenides ($+12.5$) and ternary 3$d$ borides,
carbides and silicides ($+11.3$); the largest advantages are plain ternary
intermetallics ($-7.7$) and ternary lanthanide intermetallics ($-6.6$).

\begin{table}[htbp]\centering
\caption{Matched cells, GNoME against Materials Project previously catalogued materials;
the ten of twenty whose intervals exclude zero. $n_{\text{G}}$ and $n_{\text{K}}$ are the
GNoME and known-material counts; the GNoME rate is release-based and the known rate is a
confirmed transition (Section~\ref{sec:defs}). Positive $\delta$ means GNoME disagrees
more; $\delta$ is computed from unrounded rates. Intervals are percentile intervals from
20{,}000 replicates: the GNoME side is a
cluster bootstrap over structural prototypes drawn across all twenty cells at once, the
Materials Project side a material-level bootstrap. No multiplicity correction is applied
and the comparison is exploratory.}
\label{tab:known}
\small\begin{tabular}{lrrrrrc}
\toprule
Cell & $n_{\text{G}}$ & GNoME & $n_{\text{K}}$ & Known & $\delta$ & 95\% CI \\
\midrule
4-el oxide, 3$d$ & 2{,}478 & 30.7\% & 180 & 13.3\% & $+17.3$ & [$+11.4$, $+23.0$] \\
3-el chalcogenide, 3$d$ & 564 & 30.3\% & 450 & 17.8\% & $+12.5$ & [$+6.0$, $+19.4$] \\
3-el bor./carb./sil., 3$d$ & 575 & 29.6\% & 454 & 18.3\% & $+11.3$ & [$+4.6$, $+18.0$] \\
3-el halide, no Ln/3$d$ & 1{,}326 & 17.3\% & 1{,}125 & 9.2\% & $+8.2$ & [$+4.7$, $+11.9$] \\
4-el oxide, no Ln/3$d$ & 3{,}318 & 18.9\% & 297 & 10.8\% & $+8.2$ & [$+3.9$, $+12.4$] \\
3-el oxide, no Ln/3$d$ & 921 & 21.7\% & 1{,}553 & 16.7\% & $+5.0$ & [$+1.4$, $+8.6$] \\
3-el pnictide, no Ln/3$d$ & 1{,}473 & 11.1\% & 852 & 6.9\% & $+4.1$ & [$+1.3$, $+7.1$] \\
3-el chalcogenide, no Ln/3$d$ & 1{,}106 & 10.6\% & 1{,}064 & 7.4\% & $+3.2$ & [$+0.5$, $+6.0$] \\
3-el intermetallic, Ln & 4{,}661 & 6.4\% & 1{,}988 & 13.0\% & $-6.6$ & [$-8.7$, $-4.5$] \\
3-el intermetallic, no Ln/3$d$ & 2{,}618 & 7.9\% & 1{,}144 & 15.7\% & $-7.7$ & [$-10.2$, $-5.3$] \\
\bottomrule
\end{tabular}
\end{table}

The aggregate figures and the matched cells run in opposite directions, and the
difference comes from which cells are compared. Within the twenty matched cells GNoME
disagrees \emph{more}, 15.50\% against 14.01\%; standardised to the known-materials
cell mix the difference is $+1.23$ points [$+0.22$, $+2.28$] and to the GNoME mix
$+1.85$ [$+0.57$, $+3.11$], both excluding zero (bootstrap means; the plug-in
standardised differences are $+1.22$ and $+1.84$). What differs in sign is the comparison
between the \emph{full} populations: the full release is 9.8\% oxide where the
matched-cell GNoME population is 33.3\% oxide, so restricting to cells the known
materials also populate upweights oxides roughly threefold. That is a descriptive
decomposition of two rates, not a demonstration that composition \emph{causes} the
difference. The Materials Project side is a confirmed transition while the GNoME side
above is release-based; re-evaluating the GNoME side on the confirmed-transition basis
leaves seventeen cells (three fall below 500 GNoME materials on that basis) with
standardised differences of $+1.33$ [$+0.18$, $+2.51$] and
$+2.25$ [$+0.83$, $+3.66$], the same answer slightly more strongly.

An incomplete
r$^2$SCAN hull omits competing phases and so understates flip rates on either side. On
the Materials Project side, the r$^2$SCAN hulls of known oxides are the least complete of
any anion class (median 37\% of competing GGA entries recomputed) and those of
intermetallics the most complete among the large classes (64\%,
Section~\ref{sec:allref}); completing the Materials Project side alone could shrink the
$+17$-point gap in 3$d$ oxides or leave it unchanged, so the $+17$ is an upper figure only in the weak sense that this one-sided operation
cannot exceed it, and not a bound on a comparison in which
both sides are completed. In intermetallics, GNoME's own hulls gain most in proportional terms from the denser
all-entries reference,
so part of its $-7.7$-point advantage there may be incompleteness on the GNoME side.
The comparison is also not matched on distance below the PBE hull, the single covariate
most likely to confound it; the Materials Project thermo documents expose
\texttt{decomposition\_enthalpy}, the quantity defined by \citet{bartel2019}, so the omission is ours; we leave it to a
follow-up rather than add an unregistered analysis at this stage. The defensible claim is therefore:
\emph{within the analysed datasets, functional-level disagreement differs between GNoME
and previously catalogued materials in a chemistry-dependent manner, and the direction
of the aggregate comparison differs from that seen in several matched composition
cells.} It is not a claim that the mode of discovery as such makes a material's label
more or less reliable.

\subsection{Verification budgets: composition adds to hull distance}\label{sec:budget}

If flip rates are predictable from composition, a maintainer with a limited r$^2$SCAN
budget should not sample uniformly. Uniform sampling, however, is a weak thing to beat,
and there is an obvious competitor that costs nothing: the PBE decomposition energy
itself, already known for every candidate. Section~\ref{sec:hulldist} shows that flips
concentrate steeply toward the hull. Any claim for composition-aware allocation has to be
made against that baseline, so we evaluate four policies.

The population for the budget evaluation is the 117{,}018 recomputed materials that
carry a PBE energy and were PBE-stable, in 7{,}822 prototype clusters, half of which form the estimation
set and the rest the evaluation set. \textbf{The split is by cluster, not by
material}. The composition policy estimates a flip rate for each (element
count, anion class, lanthanide, 3$d$) cell from the estimation set only and ranks
evaluation materials by it, ties broken by a seeded permutation; the hull-distance
policy ranks by PBE decomposition energy, closest first, with no estimation step; the
combined policy ranks by the fitted probability from a logistic regression of the flip
outcome on the estimated cell rate and the hull distance, fitted on the estimation set.
A budget $b$ selects the first $\lfloor bN \rfloor$ of the $N$ evaluation materials,
and the \emph{catch} is the share of the evaluation set's flips that the selection
surfaces; uniform random selection has expected catch $b$. The procedure is repeated over
20 cluster-wise splits (full statement in Appendix~\ref{app:impl}); the across-split
spread describes how much the answer moves with the split and is not a confidence
interval, and the budget counts entries, not compute.

\begin{table}[htbp]\centering
\caption{Share of flipped labels surfaced at a given recomputation budget, under four
ranking policies. Mean over 20 cluster-wise splits; standard deviation across splits in
brackets; the uniform-random column is the analytic expectation and carries none; the
best policy at each budget is in bold. Hull distance alone outperforms composition alone
at every budget; the two together outperform either.}
\label{tab:budget}
\small\begin{tabular}{lrrrr}
\toprule
Budget & Uniform random & Composition & Hull distance & Both \\
\midrule
\phantom{0}5\% & \phantom{0}5.0\% & 11.2\% [0.5] & 12.9\% [0.3] & \textbf{14.3\%} [0.5] \\
10\% & 10.0\% & 20.8\% [0.8] & 23.0\% [0.7] & \textbf{26.5\%} [0.9] \\
20\% & 20.0\% & 37.6\% [1.0] & 40.4\% [0.9] & \textbf{45.8\%} [1.1] \\
30\% & 30.0\% & 50.0\% [0.9] & 54.8\% [0.8] & \textbf{61.6\%} [1.0] \\
50\% & 50.0\% & 71.0\% [1.0] & 77.0\% [0.8] & \textbf{84.0\%} [0.6] \\
\bottomrule
\end{tabular}
\end{table}

\begin{figure}[htbp]\centering
\includegraphics[width=0.71\textwidth]{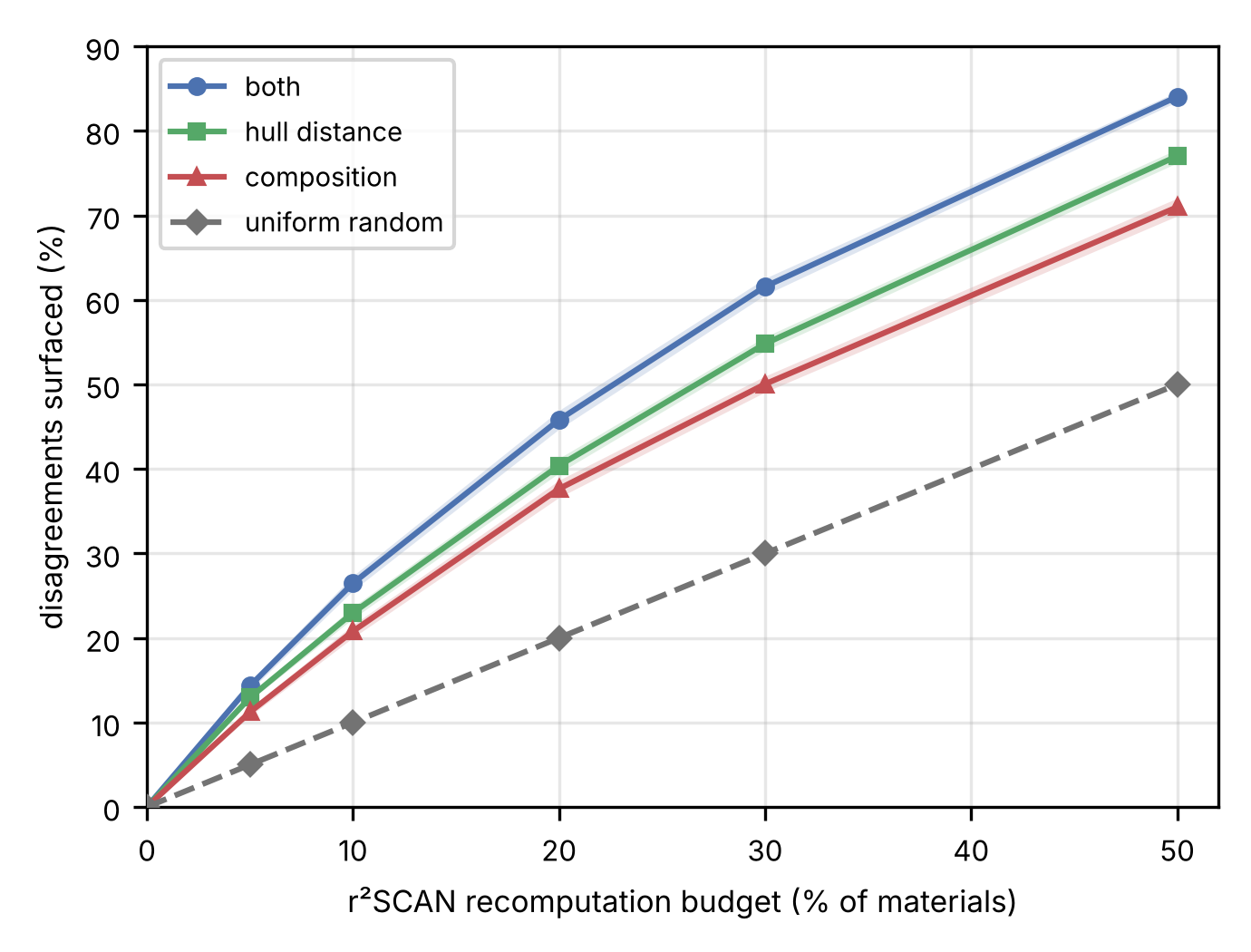}
\caption{Disagreements surfaced against recomputation budget under four ranking
policies, anchored at zero budget. Means over 20 cluster-wise splits; the shaded bands,
$\pm1$ standard deviation across splits, are barely wider than the lines.}
\label{fig:budget}
\end{figure}

The ordering is the result (Table~\ref{tab:budget}, Figure~\ref{fig:budget}). Ranking
by composition alone surfaces 37.6\% of disagreements at a 20\% budget against 20.0\%
for uniform sampling, but ranking by hull
distance alone does better still, at 40.4\%, and at every budget we tested. \textbf{A
maintainer who used this paper's chemistry cells in place of the PBE energy already in
their database would be worse off.} Because the policies share the same 20 splits, the
paired per-split differences are the relevant uncertainty: hull distance beats
composition by $+2.75$ points at a 20\% budget (paired s.d.\ 1.43, direction held in
20 of 20 splits) and the combined
policy beats hull distance alone by $+5.42$ points (paired s.d.\ 1.39, 20 of 20 splits). What
composition does do is add to hull distance: the two together surface 45.8\% at a 20\%
budget, and the gain persists at every budget and widens in absolute terms as the budget
grows. Closest-first ranking uses the variable without fitting anything, so composition might
only be standing in for a better-modelled hull distance. On 20 cluster-wise splits drawn afresh (the first coincides with the first split above) we therefore
added a hull-distance-only logistic model over twenty quantile bins of PBE depth, and a
cross-fitted combined model in which the estimation half is split five ways by prototype
cluster. The twenty-bin
hull model did not improve held-out catch over closest-first ranking ($-0.06$ points at
a 20\% budget); against it the combined policy is ahead by $+6.00$ points (paired s.d.\
0.80, 20 of 20) and the cross-fitted combined policy by $+6.31$ (paired s.d.\ 0.78, 20
of 20). Composition still adds after hull distance is modelled nonparametrically, and
the gain persists under cross-fitting, which is evidence against generated-regressor
optimism rather than proof that none is present. We claim neither that the combined
policy is optimal nor that its per-entry probabilities are calibrated: the evaluation
measures
held-out ranking performance.

\subsection{Robustness}\label{sec:robust}

\paragraph{Threshold.} A material counts as disagreeing when
$E_{\text{dec}}(\text{r}^2\text{SCAN}) > \tau$, with $\tau=0$ in the rest of the paper,
on the release-based population of Section~\ref{sec:defs}. We recomputed rates for $\tau$ from 0 to 50 meV/atom in 5 meV/atom steps
(Appendix~\ref{app:threshold}). Oxides rank first and the intermetallic class last at
every threshold, which is the contrast the paper's claims rest on; the middle of the
ordering reshuffles, with chalcogenides falling from fourth to sixth of the seven classes
between 0 and 5 meV/atom. Reporting against threshold rather than at a single cut follows
\citet{reents2026}.

\paragraph{Reference set.} The contrasts we tested hold against both the MP and
MP+OQMD hulls; across the 40 primary cells the MP+OQMD rate runs 1.1 to 5.1 points higher
(median 2.4), and the cell ordering is nearly preserved (Spearman $\rho = 0.99$; 30 of
780 pairwise orderings invert, all between cells whose intervals overlap). The ``All'' reference set behaves differently
and is treated in Section~\ref{sec:allref}.

\paragraph{Pseudopotentials.} GNoME used newer PAW potentials for Li, Na, Mg, Ge and Ga
\citep{merchant2023}, and its MP-referenced hull applies per-atom corrections to Materials Project entries
containing those elements \citep{gnomedataset}. Materials containing them flip
\emph{less} (11.0\% against 13.5\%). That is a
marginal association over compositions differing in much more than the potential used,
and it does not exclude a pseudopotential contribution, which would require a
within-composition comparison we have not run.

\subsection{Hull completeness, probed on both sides}\label{sec:allref}

The release documents its ``All'' column as the distance to the convex hull of all
entries \citep{gnomedataset}, which we read as GNoME's own structures alongside the
Materials Project, OQMD and WBM entries (Section~\ref{sec:datasets}). Rates against it are measured against a denser
reference, and the global flip rate is 25.8\%, roughly double the MP-referenced figure.
A more complete hull can only raise decomposition energies, and ``All'' raises them
everywhere. The chemistry that gains the most in proportional terms is intermetallics
(Table~\ref{tab:allref}), whose rate rises from 9.2\% to 22.3\%, a factor of 2.4 --- and
in the headline cell from 4.0\% to 14.5\%; in percentage points the boride, carbide and
silicide class gains slightly more (15.0\% to 29.6\%). That is consistent with the
MP-referenced hull being least complete for intermetallics, the class to which GNoME
added the largest number of new phases, but it is not a measurement of completeness: a denser hull does not identify
\emph{which} phases were added to \emph{which} chemistry, and a larger change in a
binary flip rate can also reflect how many materials sit near the classification
boundary, so the observed quantity is the rate's sensitivity to the reference-set
definition.

\begin{table}[htbp]\centering
\caption{Anion-class flip rate against the MP reference hull and the ``All'' reference set.}
\label{tab:allref}
\small\begin{tabular}{lrrr}
\toprule
Anion class & $n$ & MP reference & ``All'' reference \\
\midrule
oxide                   & 13{,}668 & 19.3\% & 29.5\% \\
pnictide                & 26{,}330 & 16.7\% & 29.6\% \\
boride/carbide/silicide & 25{,}732 & 15.0\% & 29.6\% \\
chalcogenide            & \phantom{0}7{,}849 & 12.1\% & 21.6\% \\
hydride                 & \phantom{0}1{,}827 & 12.0\% & 22.3\% \\
halide                  & \phantom{0}9{,}840 & 11.5\% & 24.2\% \\
intermetallic           & 54{,}295 & \phantom{0}9.2\% & 22.3\% \\
\bottomrule
\end{tabular}
\end{table}

The gradient survives, but in a weaker form than the threshold check of
Section~\ref{sec:robust} found --- oxides drop from first to third among the three
highest classes (29.5\% against 29.6\% and 29.6\%) and intermetallics are no longer the
lowest class (22.3\% against 21.6\% for chalcogenides) --- while the
oxide-to-intermetallic ratio compresses from 2.1-fold under the MP reference to 1.3-fold
under ``All''. We report these as estimates under two reference-set definitions, not as
bounds, since two observed choices do not bracket the chemistry effect; what the
comparison establishes is that the effect is substantially reference-set sensitive. We
kept the MP reference as primary because it reproduces the source paper's published aggregate,
not because it gives the larger spread.

Table~\ref{tab:allcells} answers the narrower question on the two cells the abstract
quotes. Both rates rise under ``All'' --- from 30.7\% to 37.4\% for lanthanide-free
3$d$-containing quaternary oxides, and from 4.0\% to 14.5\% for lanthanide-free 3$d$-free
quaternary intermetallics --- so the ratio compresses from 7.67-fold to 2.58-fold while
the absolute spread falls only from $+26.7$ to $+22.9$ points, with an interval of
[$+19.7$, $+26.3$] far from zero. The headline contrast survives the alternative
reference set; what does not survive is its magnitude as a ratio.

\begin{table}[htbp]\centering
\caption{The two headline composition cells under both reference-set definitions. The
``All'' intervals and both spread and ratio intervals are cluster bootstraps over
structural prototypes at 20{,}000 replicates, with one prototype draw per replicate
shared across the two cells. The MP-reference cell rates are given as point estimates
only; their intervals are the per-cell draws of Appendix~\ref{app:cells}.}
\label{tab:allcells}
\small\begin{tabular}{lrrr}
\toprule
 & $n$ & MP reference & ``All'' reference \\
\midrule
oxide, no Ln, 3$d$ (4-el)        & 2{,}478 & 30.7\% & 37.4\% [34.6, 40.4] \\
intermetallic, no Ln/3$d$ (4-el) & 5{,}025 & \phantom{0}4.0\% & 14.5\% [12.9, 16.1] \\
\midrule
spread (pp)                      &         & $+26.7$ [$+23.5$, $+29.9$] & $+22.9$ [$+19.7$, $+26.3$] \\
ratio                            &         & 7.67 [6.29, 9.52] & 2.58 [2.26, 2.97] \\
\bottomrule
\end{tabular}
\end{table}

\paragraph{The Materials Project side can be inspected directly.} For each of the
22{,}756 known materials in Section~\ref{sec:known} we define r$^2$SCAN hull completeness
as the fraction of GGA(+$U$) entries in the material's chemical system and all of its
subsystems, excluding the material itself, that also carry an r$^2$SCAN thermo document.
This counts every entry rather than the phases that define the hull, but it is a direct
count, and it varies strongly with chemistry (Table~\ref{tab:mpcompl}): oxides have the
lowest median coverage of any class and intermetallics the highest among the large
classes, opposite to what the ``All'' comparison would suggest for GNoME's own hulls. When the known materials are split into completeness
terciles (cut at 46.2\% and 61.1\%), the flip rate for materials containing a 3$d$
element is 21.5\%, 19.8\% and 18.7\% against 13.4\%, 9.9\% and 13.4\% for materials
without one, an excess of $+8.2$, $+9.9$ and $+5.3$ points that remains above $+5$ in the most
complete tercile; within the lowest tercile, the only one holding more than a handful of
3$d$ oxides (1{,}205, against 9 and 0 in the other two), they flip at 21.0\% against
14.6\% for plain intermetallics ($n=315$). Stratifying on the coverage we can count does not remove the
3$d$-associated contrast; it does not establish that the contrast is free of a
completeness artefact.

\begin{table}[htbp]\centering
\caption{Materials Project side: median r$^2$SCAN hull completeness and flip rate by
anion class, for the 22{,}756 GGA(+$U$)-stable materials with r$^2$SCAN data.}
\label{tab:mpcompl}
\small\begin{tabular}{lrrr}
\toprule
Anion class & $n$ & Median completeness & Flip rate \\
\midrule
oxide                   & 4{,}065 & 37.0\% & 17.0\% \\
intermetallic           & 7{,}954 & 64.1\% & 16.4\% \\
pnictide                & 2{,}801 & 53.3\% & 15.2\% \\
halide                  & 2{,}634 & 58.2\% & 13.0\% \\
boride/carbide/silicide & 2{,}365 & 42.0\% & 12.1\% \\
chalcogenide            & 2{,}704 & 58.5\% & 11.7\% \\
hydride                 & \phantom{0,}233 & 67.3\% & \phantom{0}9.4\% \\
\bottomrule
\end{tabular}
\end{table}

\section{Selection effects}\label{sec:selection}

Beyond the propensity analysis of Section~\ref{sec:ipw}, two coverage features constrain
what can be claimed.

\paragraph{Element count.} Coverage is 12.9\% for binaries, 31.8\% for ternaries and
31.4\% for quaternaries, then collapses to 0.13\% for quinaries (229 of 177{,}392) and
zero for senaries (0 of 4{,}792). GNoME highlights discovery in combinatorial spaces of five or more elements
\citep{merchant2023}, and quinaries make up 32\% of the released dataset, yet
the recomputed coverage there is insufficient to support the stratified claims we make
for ternaries and quaternaries. Our logistic model finds disagreement risk
\emph{decreasing} with element count, but extrapolating a fitted trend past the edge of
the data is exactly the move this paper argues against.

\paragraph{Distance to the hull.} Over all 554{,}054 summary entries, coverage peaks at 26.1\% in the
$(-0.1,-0.05]$ eV/atom band (it is 18.3\% below $-0.2$ and 25.8\% in $(-0.2,-0.1]$) and declines
monotonically from
there toward the hull, to 16.5\% in $(-0.01,0]$ and 0.3\% above zero. These band
coverages pool all element counts, so part of the gradient is the element-count effect
just described; the propensity model of Section~\ref{sec:ipw} enters both. Either way,
the recomputed subset under-samples precisely the region where flips are most likely.
Taken alone that would make the observed rate an underestimate of the population rate,
but the propensity model includes PBE decomposition energy alongside
composition and cell size, and the net effect of weighting is to move the estimate
\emph{down}, from 13.36\% to 12.71\%. We did not decompose that net change into
per-feature contributions, and we claim no direction for the overall selection bias.

\paragraph{Censoring bounds.} Of the $N=365{,}504$ summary-file entries that are
PBE-stable and have at most four elements, $n=116{,}789$ were recomputed, among which
$F=15{,}608$ flipped. Treating every unrecomputed entry among the $N$ as unflipped gives
$F/N = 4.3\%$; treating every one as flipped gives $(F + (N-n))/N = 72.3\%$. These bounds
do not identify a useful population rate, as any worst-case bound on 68\% censoring
cannot. The 12.7\% propensity-weighted result is reported separately as a
model-dependent sensitivity analysis, not as a replacement for those bounds.

\paragraph{Unmatched identifiers.} A total of 22{,}498 r$^2$SCAN material IDs do not appear in the
current summary file. Their distribution over anion class and element count is close to
that of the matched remainder, and their flip rate is 2.0 points lower (11.4\% against
13.4\%), which does not suggest selective omission of high-disagreement entries. One
element is affected wholesale: every ytterbium-containing entry in the r$^2$SCAN file
(6{,}265 of them, 28\% of the unmatched set) lacks a summary-file row, so no PBE-matched
analysis contains ytterbium. We state a
release-version mismatch as a hypothesis (Appendix~\ref{app:manifest}). Including or
excluding them changes no conclusion.

\section{Discussion}

\paragraph{For users of generated catalogues.} A single aggregate reliability figure is
insufficient for chemistry-specific prioritisation. A group screening GNoME oxides faces
a 19.3\% rate of verdicts that r$^2$SCAN would overturn against 9.2\% for
intermetallics, and narrowing to the cell widens the gap further: quaternary
lanthanide-free 3$d$ oxides flip at 30.7\% against 4.0\% for the corresponding plain
intermetallics. We suggest datasets of this kind ship a per-entry verification-priority score rather
than a global statistic, combining proximity to the hull with composition, in that order
of importance (Section~\ref{sec:budget}); what is demonstrated here is ranking
performance, and calibrated per-entry probabilities would be a separate objective. The rates above are those of the analysed
GNoME cells and should not be carried over to other databases or to benchmark test sets.

\paragraph{For benchmark design.} Benchmarks such as Matbench Discovery evaluate MLIPs
against a PBE hull \citep{riebesell2025}, correctly for their purpose of measuring
whether a model reproduces DFT. If a model is scored on how well it reproduces PBE
verdicts, the chemistry composition of the test set determines how much of the score
reflects agreement with labels a higher-fidelity functional would overturn: within the
GNoME cells we analyse, roughly 30\% for lanthanide-free 3$d$ quaternary oxides and
roughly 4\% for the corresponding plain intermetallics. We make no claim about any
particular benchmark result.

\paragraph{For MLIP training data.} Warford et al.\ argue that selectively applied $U$
splits a training set into incompatible surfaces \citep{warford2026}.
Section~\ref{sec:u} shows that label disagreement is associated with the element set
they identify, though our comparisons do not establish the correction as its cause.
Warford et al.\ report that datasets omitting $+U$ entirely, such as MatPES
\citep{kaplan2025} and MP-ALOE \citep{kuner2025}, avoid the issue \citep{warford2026};
nothing here tests that.

\paragraph{What would settle the physics.} Our evidence is statistical: $U$-list
content predicts disagreement; the 3$d$-associated excess persists within band-gap
bins, the cell contrast within hull-distance bands, the cell contrast under inverse-probability weighting for recomputation selection,
and the $U$-list association under adjustment for cell size and across the $U$ boundary
itself. We do not show \emph{why} the two functionals
disagree for these materials; that requires a targeted study of a small number of
systems at higher levels of theory.

\section{Limitations}\label{sec:limits}

\paragraph{Hull completeness is the largest partially resolved threat.} An r$^2$SCAN
convex hull omits any competing phase for which no r$^2$SCAN calculation exists,
\emph{understating} flip rates; the pure r$^2$SCAN thermo type we query does not guard
against this as the Materials Project's mixed scheme does \citep{mpmixingdocs}, and the
GNoME decomposition energies were computed against hulls we cannot inspect. If
reference-phase coverage varies systematically with chemistry, part of the gradient
could reflect differential hull completeness. Section~\ref{sec:allref} is a partial
probe: under the ``All'' reference set the oxide-to-intermetallic ratio falls from 2.1
to 1.3 while oxides remain among the three highest classes, and the 3$d$-associated contrast
persists within each tested entry-coverage stratum on the Materials Project side. A
direct test on the GNoME side would require the dataset authors' hull construction,
which is not released.

\paragraph{Neither functional is ground truth.} r$^2$SCAN is more accurate than PBE for
formation energies on average \citep{kothakonda2023}, and than SCAN and PBEsol
\citep{kingsbury2022prm}, but not uniformly: for transition metals and intermetallics
PBE and r$^2$SCAN are comparable \citep{kothakonda2023}, PBE and SCAN perform similarly
on decomposition reactions \citep{bartel2019}, and for copper SCAN and r$^2$SCAN
over-stabilise Cu$^{2+}$ relative to Cu$^{+}$, an error that PBE+$U$ avoids but that a $U$
added to SCAN does not repair \citep{bae2025}; the Materials Project applies no $U$ to
copper, so GNoME's copper labels carry no $U$ and neither workflow benefits from that repair. A low flip rate for a chemistry says only that
the two functionals agree there.

\paragraph{Clustering is corrected but the proxy is imperfect.} Our prototype proxy
will merge genuinely distinct structures sharing a space group and stoichiometry pattern
and split related structures that relaxed into different space groups, and because
compositions are not reduced, the same compound released as A$_2$B$_4$ and as AB$_2$
would receive different patterns. A true prototype assignment via structure matching
would be better. The headline result is unchanged under a coarser definition and under
removal of the largest clusters (Section~\ref{sec:cluster}), and we claim no direction
for the residual error.

\paragraph{The model supports ranking, not per-material prediction.} The McFadden
pseudo-$R^2$ is 0.066, and the held-out evaluation of Section~\ref{sec:budget} supports
the model's ability to \emph{rank} materials rather than to assign calibrated per-entry
probabilities. We have not assessed calibration.

\paragraph{Selection of the candidates themselves is not resolved.}\label{sec:curse}
GNoME's candidates were filtered for PBE stability by a PBE-trained model, so the
released set may be enriched for structures where PBE happens to be favourable --- a
winner's curse that would predict elevated flip rates with no chemistry-functional
interaction required. The release supports two internal probes, the generation pipeline
(\texttt{Data Directory}) and training-set membership (\texttt{Is Train}): using
lanthanide-pooled cells, oxide+3$d$ minus 3$d$-free intermetallic is $+20.0$ and $+22.3$
points across the pipeline split and $+21.6$ and $+19.4$ across the training-set split.
Four contrasts spanning 2.9 points, with no intervals, make a weak test, and it does not
exclude a shared chemistry-dependent selection process, since a common energy
displacement need not produce equal changes in flip probability across classes sitting
at different distances from the hull.

\paragraph{The matched comparison is the weakest step.} Four-element known cells are
small ($n=108$ to $297$), ``known'' means previously catalogued including theoretical
entries, the comparison is unmatched on hull distance (Section~\ref{sec:known}), and it is
the comparison most exposed to selection on the candidates themselves. We treat it as
exploratory throughout.

\paragraph{Partly unexplained features.} Nickel's position at the global average
despite receiving $U$ is consistent with its nearly filled spin-up $d$ bands limiting any meta-GGA moment
enhancement (Section~\ref{sec:u}); the reversal of its anion-class ordering
relative to iron and cobalt is not accounted for by that argument, or by any other we can
test here. The triclinic excess is unexplained.

\paragraph{Scope.} The GNoME results apply to materials with at most four elements that
were selected for r$^2$SCAN recomputation; the 266 recomputed quinaries (229 of them matched to the summary file) are included in
the global rate but support no stratified claim. Sections~\ref{sec:known} and~\ref{sec:allref} additionally report quantities computed on Materials Project entries,
which are drawn from a different population and are labelled as such throughout.

\section{Conclusion}

A single aggregate survival rate is not an adequate description of how far a generated
catalogue's stability labels depend on the workflow that produced them. Resolved by
chemistry, the verdict-flip rate between GNoME's PBE(+$U$) labels and its r$^2$SCAN
recomputations ranges from 4.0\% [3.3, 4.7] for plain quaternary intermetallics to
30.7\% [27.7, 33.8] for lanthanide-free 3$d$-containing quaternary oxides. The contrast
persists under correction for non-independence among substitution-generated structures
and under inverse-probability weighting for recomputation selection; a
lanthanide-pooled form of it is present within every populated hull-distance band and
every generation-pipeline and training-set split examined, and the 3$d$-associated excess
within every band-gap bin; its magnitude is
substantially sensitive to the reference-set definition, the oxide-to-intermetallic
ratio running from 2.1 to 1.3 across the two sets we can compare.

An additive per-element account is not what the data look like: a composition-linear
offset cancels from a decomposition energy by construction, and in a fitted model that
allows context dependence an element's coefficient varies strongly with what it is
combined with. Within a post-selected element set, cluster-robust joint tests reject
equal context coefficients for iron, chromium, manganese and nickel and not for
cobalt. Underlying the binary verdict is a chemistry-dependent
difference in mean displacement on top of much larger scatter: $+14.7$ meV/atom for 3$d$
oxides against $-8.9$ for plain intermetallics, with within-cell standard deviations of 24 to 56
meV/atom. Membership of the Materials Project Hubbard-$U$ list is a strong observed
compositional predictor, and selective application of $U$ is not a sufficient
explanation of the pattern: elevated disagreement persists outside the $U$-treated
regime, and the highest within-class iron rates are in borides, pnictides and
intermetallics, chemistry receiving no $U$. The smaller Fe-associated excess in oxides is consistent
with partial suppression by the corrected treatment but does not establish it; for cobalt
the reading is ambiguous, and nickel runs the other way. We do not test a physical mechanism; a plausible candidate is
meta-GGA over-magnetisation of itinerant 3$d$ systems. Lanthanide content is associated
with a below-average rate overall, as the pre-committed claim anticipated (a claim
informed by an earlier look at the same data), but the association reverses sign
with the anion and is near zero when averaged over classes because within-class effects
of opposite sign cancel.

For verification budgets, composition supplements proximity to the hull rather than
replacing it. In cluster-held-out evaluation at a 20\% entry-count budget, ranking by
PBE hull distance alone surfaces 40.4\% of disagreements and the combined policy
45.8\%, against 37.6\% for composition alone; the combined policy stays about six points
ahead of a twenty-bin nonparametric model of hull distance, and cross-fitting the
estimated cell rates does not remove the gain. That is a held-out ranking result and not
a claim about calibrated per-entry risk.

These findings concern sensitivity to computational workflow, not experimental
accuracy: the released outcomes do not permit separate attribution of disagreement to
functional choice, relaxation, empirical corrections and reference-phase coverage. The
matched comparison against previously catalogued materials is exploratory, and coverage
beyond four elements --- the regime GNoME advertises most prominently --- is
insufficient for the stratified claims made here.

\section*{Data and code availability}

All input data are public: the two GNoME CSV files are served from the DeepMind bucket
\texttt{gs://gdm\_materials\_discovery/gnome\_data/} (139{,}541 and 554{,}054 rows at the
time of access, 9 October 2026), and the Materials Project thermo records were retrieved
through the public API on 9--10 October 2026. Appendix~\ref{app:manifest} gives the full
input manifest. The repository at \url{https://github.com/FOUNDERACCRETION/gnome-audit}
contains the pre-committed decision rules with their commit history, one script per
analysis phase, every machine-readable table and figure the paper draws on, pinned
package versions, and a \texttt{run\_all.py} driver that regenerates the results
directory from the two CSV files and then checks the regenerated output against every
number printed here: eighty-four headline quantities, the values parsed out of the
\LaTeX{} tables, and the prose. A continuous-integration workflow performs the whole
regeneration on a clean runner on every push that touches the analysis code, after
deleting the stored results so that the checks cannot pass on them.
\texttt{VALIDATION.md} states what each check asserts and what none of them can
establish. The only components not reproducible without credentials are the two
Materials Project pulls, which require a free API key; their cached outputs are
committed, and the Materials Project changes over time: a live re-pull on 22 October
2026 returned two additional boride/carbide/silicide entries and moved eight matched-cell
interval endpoints by at most 0.2 points, with every point estimate unchanged. Reproducing a number is not the same as the number being right: these checks
test consistency between code, results and manuscript, and they will pass on a
statistical choice that is internally consistent and still wrong. \texttt{REVIEWER.md}
names the definitions, encodings, contrasts and the propensity model as the places most
worth doubting. At the time of writing the repository has not been deposited in a DOI-minting
archive.

\section*{Disclosures}

AI assistance. Claude (Anthropic) was used to assist with implementation of the analysis
code and preparation of manuscript text. I independently defined the research question and
scope, obtained the data, specified and committed the decision rules before the stratified
analysis, executed the Materials Project queries, reviewed and tested the code and outputs,
and revised the final manuscript. The verification actually performed is described above, under
Data and code availability, and in the next paragraph, rather than asserted here. All
analytical decisions, interpretations and claims are my responsibility. The complete code,
machine-readable outputs, validation checks and reproducibility workflow are publicly
available.

Every reference has been checked against the publisher, journal or arXiv record. The following
numerical and interpretive claims were additionally checked against the source text rather
than against the citation alone: Merchant et al.'s 84\%, 86.8\% and 90\% figures, their
381{,}000 and 421{,}000 counts and the 232{,}477 attributed to symmetry-aware partial
substitution \citep{merchant2023}; the Materials Project correction scheme's fourteen
corrected anion species and its restriction of the Hubbard $U$ to eight transition metals
in oxides and fluorides \citep{mpanion}; the 6{,}307 materials behind Kingsbury et al.'s ``approximately 6000
solid materials'' and their conclusion that r$^2$SCAN predicts formation energies more accurately
than both SCAN and PBEsol \citep{kingsbury2022prm}; Kothakonda et al.'s conclusion that
GGAs are comparable to meta-GGAs for transition metals, intermetallics and weakly bound
solids, and that PBE is slightly more accurate for the 233 Type 2 decomposition enthalpies
\citep{kothakonda2023}; An et al.'s 603-to-298, 166 and 109 counts, the use of SCAN rather
than r$^2$SCAN for their refinement, and the exclusion of 142 rare-earth compounds for
reasons of $f$-electron and pseudopotential uncertainty \citep{an2026}; Yu et al.'s 19{,}998
unary and binary materials and their per-element formation-energy error metric
\citep{yu2024}; and Huang et al.'s 380{,}190 paired MatPES entries together with their
thermodynamic validation on six binary systems, among them Fe--O, Mn--O, Ti--Cl, Mo--Cl
and V--S \citep{huang2026}. The ``over 60\%'' metallic figure of \citet{liu2024} is cited as reported and is not used in any analysis; the anion-class mix
is measured directly in this paper (Table~\ref{tab:allref}). The remaining references were
verified for existence, authorship and venue but not claim by claim.

Competing interests: I am the founder of Accretion Labs, a materials-informatics company.

\appendix

\section{Cell-level results with cluster bootstrap intervals}\label{app:cells}

This appendix lists all disjoint cells with $n \ge 500$. Intervals resample prototype
clusters. The final column is the ratio of the cluster interval width to the naive Wald binomial width.

\footnotesize
\begin{longtable}{llccrrrr}
\caption{Flip rate by composition cell, all disjoint cells with $n \ge 500$, with cluster bootstrap 95\% intervals and the widening factor relative to the naive binomial interval.}\label{tab:appcells}\\
\toprule
$n_{\text{el}}$ & Anion & Ln & 3$d$ & $n$ & Rate & Cluster 95\% CI & Widen \\
\midrule
\endfirsthead
\toprule
$n_{\text{el}}$ & Anion & Ln & 3$d$ & $n$ & Rate & Cluster 95\% CI & Widen \\
\midrule
\endhead
4 & oxide & no & yes & 2{,}478 & 30.7\% & [27.7, 33.8] & 1.69 \\
3 & chalcogenide & no & yes & 564 & 30.3\% & [24.9, 36.5] & 1.53 \\
4 & pnictide & yes & yes & 4{,}573 & 30.1\% & [25.9, 33.8] & 2.96 \\
3 & boride/carbide/silicide & no & yes & 575 & 29.6\% & [24.1, 35.3] & 1.51 \\
4 & boride/carbide/silicide & no & yes & 4{,}501 & 25.2\% & [21.5, 29.5] & 3.14 \\
3 & pnictide & no & yes & 813 & 24.0\% & [19.9, 28.7] & 1.51 \\
4 & boride/carbide/silicide & yes & yes & 4{,}561 & 22.8\% & [18.5, 27.4] & 3.65 \\
4 & halide & no & yes & 794 & 22.4\% & [17.5, 27.9] & 1.80 \\
3 & oxide & no & no & 921 & 21.7\% & [18.7, 24.9] & 1.17 \\
3 & intermetallic & no & yes & 2{,}667 & 21.1\% & [18.4, 24.0] & 1.80 \\
3 & intermetallic & yes & yes & 2{,}056 & 19.0\% & [15.9, 22.4] & 1.91 \\
4 & oxide & no & no & 3{,}318 & 18.9\% & [16.7, 21.4] & 1.79 \\
4 & oxide & yes & yes & 1{,}399 & 18.1\% & [14.2, 22.0] & 1.94 \\
4 & pnictide & no & yes & 5{,}736 & 17.8\% & [14.7, 21.0] & 3.18 \\
3 & halide & no & no & 1{,}326 & 17.3\% & [14.3, 20.8] & 1.60 \\
3 & pnictide & yes & no & 1{,}349 & 16.8\% & [11.9, 22.4] & 2.63 \\
4 & chalcogenide & no & yes & 1{,}487 & 16.3\% & [13.4, 19.7] & 1.67 \\
4 & pnictide & yes & no & 7{,}412 & 13.9\% & [11.5, 16.5] & 3.15 \\
4 & oxide & yes & no & 4{,}232 & 13.8\% & [11.4, 16.6] & 2.52 \\
4 & halide & no & no & 1{,}404 & 13.1\% & [10.2, 16.6] & 1.82 \\
4 & intermetallic & yes & yes & 10{,}659 & 12.9\% & [11.6, 14.5] & 2.27 \\
3 & oxide & yes & no & 730 & 12.2\% & [7.7, 16.9] & 1.93 \\
4 & intermetallic & no & yes & 7{,}459 & 12.1\% & [10.7, 13.5] & 1.88 \\
3 & boride/carbide/silicide & no & no & 845 & 11.7\% & [9.0, 14.9] & 1.36 \\
3 & pnictide & no & no & 1{,}473 & 11.1\% & [8.9, 13.6] & 1.47 \\
3 & halide & yes & no & 982 & 10.6\% & [6.7, 15.8] & 2.37 \\
3 & chalcogenide & no & no & 1{,}106 & 10.6\% & [8.4, 13.0] & 1.26 \\
3 & chalcogenide & yes & no & 1{,}076 & 10.1\% & [7.4, 13.7] & 1.77 \\
3 & boride/carbide/silicide & yes & no & 1{,}246 & 9.6\% & [6.1, 13.9] & 2.37 \\
4 & halide & yes & yes & 1{,}119 & 9.1\% & [5.3, 14.5] & 2.74 \\
4 & boride/carbide/silicide & yes & no & 10{,}287 & 9.1\% & [7.0, 11.4] & 4.02 \\
4 & chalcogenide & yes & no & 1{,}497 & 8.4\% & [6.3, 10.7] & 1.59 \\
4 & boride/carbide/silicide & no & no & 3{,}422 & 8.2\% & [6.8, 9.7] & 1.59 \\
3 & intermetallic & no & no & 2{,}618 & 7.9\% & [6.6, 9.3] & 1.30 \\
4 & chalcogenide & no & no & 1{,}467 & 7.0\% & [5.5, 8.7] & 1.22 \\
3 & intermetallic & yes & no & 4{,}661 & 6.4\% & [5.0, 8.0] & 2.11 \\
4 & pnictide & no & no & 4{,}675 & 6.0\% & [4.8, 7.3] & 1.82 \\
4 & halide & yes & no & 3{,}599 & 5.9\% & [2.8, 10.8] & 5.16 \\
4 & intermetallic & yes & no & 18{,}932 & 5.5\% & [4.6, 6.3] & 2.59 \\
4 & intermetallic & no & no & 5{,}025 & 4.0\% & [3.3, 4.7] & 1.31 \\
\bottomrule
\end{longtable}
\normalsize

The median widening factor is 1.82. The highest and lowest quaternary cells remain disjoint:
30.7\% [27.7, 33.8] against 4.0\% [3.3, 4.7].

\section{Per-element results}\label{app:element}

This appendix lists all 83 elements appearing in at least 300 recomputed materials,
ordered by flip rate and read left to right across the three column groups. Entries are point estimates; cluster bootstrap
intervals for the extremes are given in Section~\ref{sec:element}. An asterisk marks the
53 elements whose contrast against the global rate survives Benjamini--Hochberg
correction at 0.05. Each element is contrasted against a global rate that includes that
element, which attenuates the contrast for the largest-$n$ entries.

\footnotesize
\begin{longtable}{lrr@{\hspace{1.6em}}lrr@{\hspace{1.6em}}lrr}
\caption{Per-element flip rate for the 83 elements with $n \ge 300$; an asterisk marks a Benjamini--Hochberg-significant contrast against the global rate.}\label{tab:appelement}\\
\toprule
El. & $n$ & Flip (\%) & El. & $n$ & Flip (\%) & El. & $n$ & Flip (\%) \\
\midrule
\endfirsthead
\toprule
El. & $n$ & Flip (\%) & El. & $n$ & Flip (\%) & El. & $n$ & Flip (\%) \\
\midrule
\endhead
Fe$^{*}$ & 5{,}762 & 42.8 & Co$^{*}$ & 10{,}972 & 33.2 & Cr$^{*}$ & 2{,}553 & 30.7 \\
Mn$^{*}$ & 4{,}994 & 25.2 & F$^{*}$ & 2{,}801 & 24.9 & As$^{*}$ & 6{,}298 & 21.6 \\
Ce$^{*}$ & 4{,}236 & 20.5 & Mo$^{*}$ & 3{,}910 & 20.2 & Sb$^{*}$ & 6{,}577 & 19.9 \\
O$^{*}$ & 13{,}668 & 19.3 & V$^{*}$ & 3{,}659 & 18.9 & Si$^{*}$ & 18{,}382 & 18.4 \\
W$^{*}$ & 3{,}223 & 18.2 & Eu$^{*}$ & 2{,}132 & 18.0 & U$^{*}$ & 2{,}595 & 17.9 \\
Te$^{*}$ & 2{,}207 & 17.4 & Bi$^{*}$ & 6{,}606 & 16.5 & Ag$^{*}$ & 4{,}224 & 16.3 \\
Gd & 976 & 16.0 & Zr & 7{,}315 & 14.7 & P & 10{,}265 & 14.2 \\
Sc & 7{,}842 & 13.9 & Se & 4{,}543 & 13.6 & Ca & 5{,}153 & 13.5 \\
Au & 5{,}932 & 13.5 & S & 3{,}383 & 13.4 & Ni & 10{,}142 & 13.4 \\
Cl & 2{,}172 & 13.2 & Y & 11{,}046 & 13.1 & Ti & 5{,}612 & 13.1 \\
Hf & 6{,}809 & 13.0 & C & 4{,}735 & 12.9 & Ga & 10{,}930 & 12.9 \\
Dy & 11{,}559 & 12.8 & K & 2{,}795 & 12.7 & H & 2{,}416 & 12.7 \\
Cu & 6{,}254 & 12.6 & Ho & 11{,}988 & 12.6 & Pr & 8{,}238 & 12.4 \\
Sr & 3{,}981 & 12.4 & Sn & 9{,}436 & 12.3 & Lu & 6{,}821 & 12.3 \\
Li & 6{,}305 & 12.2 & Ba & 3{,}580 & 12.1 & Tb$^{*}$ & 12{,}248 & 12.1 \\
Nb & 5{,}297 & 11.9 & Tm$^{*}$ & 10{,}154 & 11.8 & Er$^{*}$ & 10{,}434 & 11.6 \\
Na & 2{,}932 & 11.6 & Ta$^{*}$ & 5{,}021 & 11.3 & Nd$^{*}$ & 9{,}238 & 11.1 \\
La$^{*}$ & 6{,}546 & 11.0 & Rb$^{*}$ & 3{,}141 & 10.8 & Pu & 1{,}020 & 10.8 \\
Br & 3{,}440 & 10.6 & Pb$^{*}$ & 4{,}377 & 10.6 & B$^{*}$ & 9{,}496 & 10.6 \\
Hg$^{*}$ & 4{,}445 & 10.2 & Cs$^{*}$ & 2{,}973 & 10.1 & Th$^{*}$ & 7{,}632 & 9.9 \\
Al$^{*}$ & 11{,}933 & 9.7 & Np$^{*}$ & 1{,}387 & 9.7 & Zn$^{*}$ & 7{,}510 & 9.1 \\
Cd$^{*}$ & 5{,}356 & 9.0 & In$^{*}$ & 8{,}259 & 9.0 & Os$^{*}$ & 8{,}652 & 8.9 \\
N$^{*}$ & 3{,}845 & 8.7 & Pd$^{*}$ & 10{,}381 & 8.1 & Re$^{*}$ & 3{,}958 & 8.1 \\
Ru$^{*}$ & 12{,}381 & 8.1 & Sm$^{*}$ & 4{,}852 & 8.0 & I$^{*}$ & 3{,}091 & 7.9 \\
Pt$^{*}$ & 13{,}450 & 7.8 & Ge$^{*}$ & 4{,}764 & 7.6 & Rh$^{*}$ & 14{,}340 & 7.3 \\
Yb$^{*}$ & 6{,}265 & 7.1 & Be$^{*}$ & 2{,}799 & 6.5 & Ir$^{*}$ & 15{,}680 & 6.2 \\
Tc$^{*}$ & 7{,}350 & 5.7 & Mg$^{*}$ & 4{,}073 & 5.7 & Pm$^{*}$ & 389 & 4.4 \\
Pa$^{*}$ & 4{,}251 & 3.6 & Ac$^{*}$ & 2{,}181 & 1.4 &  & &  \\
\bottomrule
\end{longtable}
\normalsize

\section{Implementation details}\label{app:impl}

This appendix gives enough detail to re-derive the central inferences without reading
the code. Every item below is implemented in the repository script named beside it.

\paragraph{Cluster bootstrap (\texttt{phase2.py}, \texttt{phase7\_cluster.py},
\texttt{phase16\_excess\_ci.py}).} The resampling unit is the structural prototype, the
pair (space group number as recorded in the r$^2$SCAN file, anonymised stoichiometry);
compositions are used as released, not reduced, and no structure matching is performed.
Each replicate draws $G$ clusters with replacement from the $G$ clusters present in the
relevant subset and concatenates their members, so replicate sample size varies.
Intervals are percentile intervals at 2.5 and 97.5. Cell intervals and the per-element
analysis use 20{,}000 replicates; the within-anion excesses 100{,}000; the matched-cell
comparison of Section~\ref{sec:known}, 20{,}000; the coarser-clustering check, 20{,}000;
and the average marginal effects 1{,}000 refits, since each replicate there refits the
interaction model. The primary seed is \texttt{numpy.random.default\_rng(20261009)}; the average
marginal effects and the matched known-materials comparisons use 20261011, the context-spread intervals pool 20261009 and 20261012,
and the seed-stability checks below use further seeds.
Resampling is aggregated to the cluster level: drawing $G$ prototypes with replacement
has the same distribution as multinomial counts over them, so a replicate is a
matrix-vector product against per-cluster count and flip-sum vectors. Replicates in
which either arm of a contrast is empty are dropped.

The cluster pool is in every case the set of prototypes present in the subset being
resampled. Cell intervals resample the prototypes within the cell; per-element rates and the
element-versus-global contrast both come from one full-pool draw per replicate, from which
the element rate and the global rate are recomputed; within-anion excesses resample the prototypes of the anion
class and recompute both rates from the same draw. In the matched comparison
(\texttt{phase37\_matched\_cluster.py}) the GNoME side draws prototypes once per
replicate across all twenty matched cells together, and the Materials Project side is
resampled at the material level, the two sides independently; those intervals are
cluster-adjusted on the GNoME side only.

\paragraph{Monte Carlo stability.}\label{sec:mc} Replicate counts are chosen so that
the seed-to-seed movement of every endpoint we tested is below the precision it is printed
to, with two exceptions noted below (the average marginal effects and the spread
intervals of Table~\ref{tab:context}). On the lower endpoint of the Fe oxide excess, the spread across five seeds falls from
0.90 points at 400 replicates to 0.035 at 100{,}000, close to the $B^{-1/2}$ rate
(\texttt{phase33\_percentile\_conv.py}); at 100{,}000 its upper endpoint moves 0.042 points across seeds. At 20{,}000 replicates the Benjamini--Hochberg
set is the same 53 of 83 elements under all five seeds (at 300 replicates the count ranges from 51
to 53), and the
headline endpoints [27.7, 33.8] and [3.3, 4.7] come from 20{,}000 replicates at one
seed; at 600 replicates they move by up to 0.5 points across five seeds
(\texttt{phase31\_stability.py}), and the $B^{-1/2}$ scaling observed for the Fe oxide
endpoint implies movement of order 0.1 points at 20{,}000, so the last printed digit of
those endpoints carries that uncertainty. The
average marginal effects are one exception, at 1{,}000 refits: the $U$-list endpoints
move 0.03 and 0.05 points across two seeds, while the pooled lanthanide effect's upper
endpoint moves 0.10 and should be read as $+1.1$ to $+1.2$. The spread intervals of
Table~\ref{tab:context} are the other, moving by up to 13 meV/atom between seeds as its
caption states.

\paragraph{Element-versus-global contrasts (\texttt{phase8\_referee.py}).} The comparator
is estimated from the same data, so the contrast is resampled jointly: each replicate
draws one global set of clusters and recomputes both the element rate and the global
rate from it, and the standard error is that of the difference $D_e^{(b)} = \hat
p_e^{(b)} - \hat p_{\text{global}}^{(b)}$. Resampling the element rate alone against a
fixed global rate gives 51 significant contrasts after Benjamini--Hochberg rather than
53.

\paragraph{Regression encoding (\texttt{phase6b\_logit.py},
\texttt{phase10\_robust.py}).} \texttt{has\_U} is 1 if the composition contains any of
V, Cr, Mn, Fe, Co, Ni, W or Mo; \texttt{has\_3d\_nonU} is 1 only if it contains one of
Sc, Ti, Cu or Zn \emph{and} no $U$-list element; \texttt{has\_Ln} is independent of
both. The five anion dummies follow the priority rule of Section~\ref{sec:defs}, with
intermetallic and hydride pooled as the reference; element count and $\log$ sites are
$z$-scored. Coefficients are fitted by iteratively reweighted least squares; all models
converged. With 8{,}429 clusters the omitted finite-cluster factor $G/(G-1)$ is
1.0001.

\paragraph{Context-dependence model (\texttt{phase14b\_additivity.py}, identification
and inference in \texttt{phase43\_context\_identify.py}).} Table~\ref{tab:context} comes
from an ordinary-least-squares fit of $\Delta = \sum_e c_e x_e + \sum_{e \in S}\sum_{a
\in A} \gamma_{ea}\, x_e\, \mathbb{1}[\text{class}=a] + \varepsilon$ on the 117{,}018
matched, PBE-stable materials, where $x_e$ is the atomic fraction of element $e$ in the released
(unreduced) composition, the first sum runs over the 83 elements present in this subset (ytterbium is absent), $S = \{$Fe, Co, Ni, Mn,
Cr$\}$ and $A$ is the six non-hydride anion classes. There is no intercept, and the
interaction block is not reference-coded: an interaction is kept when at least 100
materials carry the element in that class, which retains all six non-hydride contexts for
every element of $S$ and drops the hydride context for each (nickel, at 99, is the
nearest), giving 30 interaction terms and 113 columns. The design is full rank (numerical rank 113 of 113 at tolerance $1.6\times10^{-9}$,
condition number $7.0\times10^{2}$), but narrowly. For each $e \in S$ the sum of its six
interaction columns equals its main-effect column on every row except hydride-class
materials, so the transformation $c_e \to c_e + t$, $\gamma_{ea} \to \gamma_{ea} - t$
for every non-hydride context $a$ leaves every non-hydride fitted value unchanged; only
the hydride rows, of which 286 contain a listed element (99 Ni, 84 Co, 62 Fe, 34 Cr,
7 Mn), anchor that direction. Deleting them drops the numerical rank to 111 with condition
number $1.8\times10^{9}$. The main-effect \emph{levels} therefore rest on a few hundred rows
and are not physical contributions. What the table reports is unaffected, because a
within-element contrast $(c_e + \gamma_{ea}) - (c_e + \gamma_{eb}) = \gamma_{ea} -
\gamma_{eb}$ is invariant under that transformation; the driver asserts that all 75 such
contrasts have inner product exactly zero with each of the five near-dependency
directions and are invariant under column scaling and permutation to
$1.6\times10^{-12}$ meV/atom.
The spread intervals are cluster bootstraps over structural prototypes, 1{,}965
replicates pooled over two seeds (965 at the first, 1{,}000 at the second; the driver
reproduces both runs), each refitting all 113 columns. The test is the
cluster-robust Wald test of Table~\ref{tab:jointtest}, on the joint null $\gamma_{ea_1} =
\cdots = \gamma_{ea_6}$ for each $e$, using the sandwich covariance over the 7{,}822
prototypes with the finite-cluster correction $\frac{G}{G-1}\cdot\frac{n-1}{n-K}$, and
Benjamini--Hochberg across the five elements. The element set $S$ was chosen after
seeing the per-element ranking, so the test is conditional on that choice.

\paragraph{Propensity weighting (\texttt{phase9\_selection.py}, diagnostics in
\texttt{phase26\_diag.py}).} The model is fitted on all 371{,}870 summary entries with at
most four elements, predicting recomputation from anion class, lanthanide and 3$d$
flags, element count, $\log$ sites and PBE decomposition energy, which enters linearly
and as its square on the standardised scale because coverage is not monotone in that
energy. Weighted outcomes are computed on the 116{,}789 recomputed, PBE-stable entries.
Over the 365{,}504 PBE-stable $\le$4-element summary entries the fitted propensity has
median 0.2995 and minimum
$e^{-500} \approx 7.1\times10^{-218}$, the underflow limit of the fit, on entries that were not recomputed; among weighted entries the
minimum is $1.52\times10^{-6}$ and the median 0.3445, against 0.2752 for the
unrecomputed. Weights are the inverse fitted probability floored at $10^{-2}$; untrimmed,
the largest weight is $6.6\times10^{5}$ and the weighted rate reads 3.05\%, an artefact
of a handful of entries. At the floor the largest weight is 100, the heaviest 1\% of
entries carry 3.3\% of the total weight, and the estimate is 12.71\%.

\paragraph{Budget evaluation (\texttt{phase12\_budget\_baselines.py},
\texttt{phase24\_budget\_strong.py}).} Eligibility is applied before splitting. Cells
absent from the estimation set receive that set's global rate; no shrinkage is applied
to small-cell estimates. The combined policy's cell rate is a generated regressor;
\texttt{phase24\_budget\_strong.py} tests that on 20 fresh splits by cross-fitting the
cell rates, dividing the estimation half into five folds by prototype cluster, and the
held-out margin over a twenty-bin nonparametric hull-distance model is $+6.31$ points
cross-fitted against $+6.00$ without cross-fitting. The uniform-random row is the analytic expectation
$\lfloor bN \rfloor / N$, not a simulation; per-split values are written to
\texttt{results/budget\_repeated.csv}.

\section{All matched cells against known materials}\label{app:allcells}

Table~\ref{tab:appallcells} lists all twenty cells of Section~\ref{sec:known}: the ten of
Table~\ref{tab:known} together with the ten whose intervals straddle zero, under the
same cell thresholds, bootstrap and conventions as that table. Positive $\delta$ means
GNoME disagrees more; $\ast$ marks the ten that exclude zero. This table,
Table~\ref{tab:known} and the standardised contrasts of Section~\ref{sec:known} are all
generated by \texttt{phase37\_matched\_cluster.py}.

\begin{table}[htbp]\centering
\caption{All twenty matched cells, GNoME against Materials Project previously catalogued materials; conventions as in Table~\ref{tab:known}; $\delta$ is computed from unrounded rates and $\ast$ marks intervals excluding zero.}\label{tab:appallcells}
\small\begin{tabular}{lrrrrrl}
\toprule
Cell & $n_{\text{G}}$ & GNoME & $n_{\text{K}}$ & Known & $\delta$ & 95\% CI \\
\midrule
4-el oxide, 3$d$ & 2{,}478 & 30.7\% & 180 & 13.3\% & $+17.3$ & [$+11.4$, $+23.0$] $\ast$ \\
3-el chalcogenide, 3$d$ & 564 & 30.3\% & 450 & 17.8\% & $+12.5$ & [$+6.0$, $+19.4$] $\ast$ \\
3-el bor./carb./sil., 3$d$ & 575 & 29.6\% & 454 & 18.3\% & $+11.3$ & [$+4.6$, $+18.0$] $\ast$ \\
3-el halide, no Ln/3$d$ & 1{,}326 & 17.3\% & 1{,}125 & 9.2\% & $+8.2$ & [$+4.7$, $+11.9$] $\ast$ \\
4-el oxide, no Ln/3$d$ & 3{,}318 & 18.9\% & 297 & 10.8\% & $+8.2$ & [$+3.9$, $+12.4$] $\ast$ \\
3-el oxide, no Ln/3$d$ & 921 & 21.7\% & 1{,}553 & 16.7\% & $+5.0$ & [$+1.4$, $+8.6$] $\ast$ \\
3-el pnictide, no Ln/3$d$ & 1{,}473 & 11.1\% & 852 & 6.9\% & $+4.1$ & [$+1.3$, $+7.1$] $\ast$ \\
3-el pnictide, 3$d$ & 813 & 24.0\% & 557 & 20.3\% & $+3.7$ & [$-1.7$, $+9.5$] \\
3-el chalcogenide, no Ln/3$d$ & 1{,}106 & 10.6\% & 1{,}064 & 7.4\% & $+3.2$ & [$+0.5$, $+6.0$] $\ast$ \\
3-el bor./carb./sil., Ln & 1{,}246 & 9.6\% & 674 & 7.1\% & $+2.5$ & [$-1.5$, $+7.2$] \\
3-el chalcogenide, Ln & 1{,}076 & 10.1\% & 442 & 7.9\% & $+2.2$ & [$-1.6$, $+6.5$] \\
3-el bor./carb./sil., no Ln/3$d$ & 845 & 11.7\% & 448 & 9.6\% & $+2.1$ & [$-1.8$, $+6.1$] \\
4-el oxide, Ln & 4{,}232 & 13.8\% & 108 & 12.0\% & $+1.8$ & [$-5.2$, $+8.2$] \\
3-el pnictide, Ln & 1{,}349 & 16.8\% & 422 & 15.4\% & $+1.4$ & [$-4.7$, $+8.0$] \\
3-el oxide, Ln & 730 & 12.2\% & 669 & 12.7\% & $-0.5$ & [$-5.8$, $+4.8$] \\
3-el intermetallic, Ln, 3$d$ & 2{,}056 & 19.0\% & 1{,}153 & 20.6\% & $-1.5$ & [$-5.4$, $+2.5$] \\
3-el intermetallic, 3$d$ & 2{,}667 & 21.1\% & 1{,}125 & 23.1\% & $-2.0$ & [$-5.7$, $+1.8$] \\
3-el halide, Ln & 982 & 10.6\% & 434 & 14.8\% & $-4.2$ & [$-9.3$, $+1.9$] \\
3-el intermetallic, Ln & 4{,}661 & 6.4\% & 1{,}988 & 13.0\% & $-6.6$ & [$-8.7$, $-4.5$] $\ast$ \\
3-el intermetallic, no Ln/3$d$ & 2{,}618 & 7.9\% & 1{,}144 & 15.7\% & $-7.7$ & [$-10.2$, $-5.3$] $\ast$ \\
\bottomrule
\end{tabular}
\end{table}

\section{Input manifest}\label{app:manifest}

\paragraph{Licensing.} GNoME data are released under the Creative Commons
Attribution-NonCommercial 4.0 International licence and the accompanying code under
Apache 2.0 \citep{gnomedataset}; they are used here for non-commercial research. This
paper's repository redistributes no GNoME records: it holds code, input checksums and
derived aggregate statistics, the largest GNoME-derived result file carrying 83 rows of
per-element summaries. The two per-material files it does commit are Materials Project
extracts, used under that project's Creative Commons Attribution 4.0 terms with the
attribution recorded in \texttt{results/PROVENANCE.md} \citep{jain2013}.

\paragraph{Files.} Two CSV files were downloaded from
\texttt{gs://gdm\_materials\_discovery/gnome\_data/} on 9 October 2026; the checksums
are of the files as downloaded, before any processing.

\noindent\texttt{stable\_materials\_summary.csv}: 554{,}054 rows, 151{,}452{,}777 bytes, SHA-256\\
\texttt{c37bffb9dd6c51d38d259a2d340ecf796194729f656d474828512390db1ff0c9}

\noindent\texttt{stable\_materials\_r2scan.csv}: 139{,}541 rows, 22{,}004{,}293 bytes, SHA-256\\
\texttt{6a811395b065f43ae7911597405eec0ecfa88251e918f4c0ba8e4e99a75656fd}

\paragraph{Columns, missing values and joins.} From the summary file we read
\texttt{MaterialId}, \texttt{Decomposition Energy Per Atom MP}, \texttt{Decomposition
Energy Per Atom All}, \texttt{Elements}, \texttt{Composition}, \texttt{NSites},
\texttt{Bandgap}, \texttt{Crystal System}, \texttt{Data Directory} and \texttt{Is Train}.
From the r$^2$SCAN file we read the same identifier, element and composition fields,
its \texttt{Decomposition Energy Per Atom MP}, \texttt{MP OQMD} and \texttt{All}
columns, and \texttt{Space Group Number}, \texttt{Crystal System} and \texttt{NSites}. Decomposition
energies are in eV/atom and band gaps in eV, as released. \texttt{Elements} is parsed
with \texttt{ast.literal\_eval} into a set; \texttt{Composition} is parsed with the
regular expression \texttt{([A-Z][a-z]?)(\textbackslash d+)} and is \emph{not} reduced.
Energy and gap fields are coerced with \texttt{pandas.to\_numeric(errors="coerce")};
rows with a missing r$^2$SCAN decomposition energy are dropped before any analysis, and
the band-gap accounting of Section~\ref{sec:gap} separates missing, infinite, zero and
positive values. The two files are joined on \texttt{MaterialId}, which is unique in both
files, so no duplicate resolution is required. The 22{,}498 r$^2$SCAN identifiers absent
from the summary file are consistent with a release-version mismatch, which we state as
a hypothesis not confirmed with the dataset authors.

\paragraph{Materials Project queries.} Thermo records were retrieved through the public
API on 9--10 October 2026 from
\url{https://api.materialsproject.org/materials/thermo/}, paging at
\texttt{\_per\_page=1000}. The \texttt{thermo\_types} parameter is passed once per pull
with the literal values \texttt{r2SCAN} and \texttt{GGA\_GGA+U} exactly as spelled; the
capitalisation is the API's and is not interchangeable, and we observed the upper-case
\texttt{R2SCAN} of the client's \texttt{ThermoType} enumeration return zero documents
without raising an error, so the spelling is worth pinning in any replication. The
\texttt{\_fields} parameter is the literal string
\texttt{material\_id,thermo\_type,energy\_above\_hull,nelements,chemsys}; no other filter
is applied. Stability is defined as $E_{\text{hull}} \le 10^{-6}$ eV/atom on each side.
Records are collapsed to one value per \texttt{material\_id}; we checked the cached
responses and found zero duplicate \texttt{(material\_id, thermo\_type)} combinations,
152{,}426 unique identifiers of 152{,}426 rows for \texttt{GGA\_GGA+U} and 37{,}123 of
37{,}123 for r$^2$SCAN. The pulls are cached in \texttt{results/mp\_gga\_raw.csv} and
\texttt{results/mp\_r2scan\_raw.csv}; they are the only components of the pipeline that
require credentials, and the stored copies are reused by the continuous-integration run.

\section{Flip rate against threshold}\label{app:threshold}

\begin{table}[htbp]\centering
\caption{Flip rate by anion class as the disagreement threshold $\tau$ is raised; B/C/Si is the boride, carbide and silicide class.}\label{tab:appthreshold}
\small
\begin{tabular}{lrrrrrrr}
\toprule
$\tau$ (eV/atom) & Overall & B/C/Si & Chalc. & Halide & Hydride & Interm. & Oxide \\
\midrule
0.000 & 13.0\% & 15.0\% & 12.1\% & 11.5\% & 12.0\% & \phantom{0}9.2\% & 19.3\% \\
0.005 & \phantom{0}9.3\% & 11.3\% & \phantom{0}8.0\% & \phantom{0}8.2\% & \phantom{0}8.6\% & \phantom{0}5.4\% & 15.5\% \\
0.010 & \phantom{0}7.1\% & \phantom{0}8.6\% & \phantom{0}6.0\% & \phantom{0}6.4\% & \phantom{0}6.0\% & \phantom{0}3.9\% & 12.5\% \\
0.020 & \phantom{0}4.5\% & \phantom{0}5.2\% & \phantom{0}3.8\% & \phantom{0}3.6\% & \phantom{0}4.3\% & \phantom{0}2.3\% & \phantom{0}8.6\% \\
0.030 & \phantom{0}3.0\% & \phantom{0}3.5\% & \phantom{0}2.6\% & \phantom{0}2.2\% & \phantom{0}2.5\% & \phantom{0}1.5\% & \phantom{0}6.1\% \\
0.040 & \phantom{0}2.1\% & \phantom{0}2.6\% & \phantom{0}2.0\% & \phantom{0}1.5\% & \phantom{0}1.7\% & \phantom{0}1.0\% & \phantom{0}4.2\% \\
0.050 & \phantom{0}1.6\% & \phantom{0}2.1\% & \phantom{0}1.6\% & \phantom{0}1.2\% & \phantom{0}1.1\% & \phantom{0}0.8\% & \phantom{0}3.2\% \\
\bottomrule
\end{tabular}
\end{table}

Table~\ref{tab:appthreshold} shows seven of the eleven thresholds computed, omitting $\tau = 15$, 25, 35
and 45 meV/atom and the pnictide column for width; the pnictide series on
the eleven-point grid $\tau = 0$, 5, 10, 15, 20, 25, 30, 35, 40, 45 and 50 meV/atom is
16.7\%, 12.8\%, 10.3\%, 8.3\%, 6.6\%, 5.2\%, 4.3\%, 3.6\%, 3.0\%, 2.5\% and 2.2\%. The
full seven-class series at all eleven thresholds is in
\texttt{results/threshold\_curves.csv}.

\end{document}